\documentclass[11pt,draftcls,onecolumn,twoside]{IEEEtran} 
\usepackage[latin1]{inputenc}
\usepackage[mathcal]{euscript}
\usepackage{amssymb,amsmath,amsthm}
\usepackage{graphicx}
\usepackage[english]{babel}
\usepackage{textcomp}
\usepackage{pstricks,pst-plot,pst-text,pst-tree,pst-eps,pst-fill,pst-node,pst-math}

\newtheorem{remark}{Remark}
\newtheorem{prop}{Proposition}

\newtheorem{problem}{Problem}
\newtheorem{theo}{Theorem}
\newtheorem{lemma}{Lemma}

\newtheorem{assumption}{Assumption}

\usepackage{color}
\definecolor{MyDarkBlue}{rgb}{0.1,0,0.65}
\definecolor{MyDarkRed}{rgb}{.85,0,0.1}

\newcommand{\EE}[1]{{\mathbb{E}}\left[ #1 \right]}

\newcommand{\dsp}{\displaystyle}
\newcommand{\dron}[2]{\frac{\partial #1}{\partial #2}}

\newcommand{\ki}{_{k,i}}
\newcommand{\ku}{_{k,1}}
\newcommand{\kd}{_{k,2}}

\newcommand{\lu}{_{l,1}}
\newcommand{\ld}{_{l,2}}

\newcommand{\K}{^{(K)}}
\newcommand{\cK}{^{c,(K)}}
\newcommand{\cpK}{^{c,(\phi_K)}}
\newcommand{\ctK}{^{c,(\theta_K)}}

\newcommand{\ApK}{^{A,(\phi_K)}}
\newcommand{\BpK}{^{B,(\phi_K)}}
\newcommand{\AK}{^{A,(K)}}
\newcommand{\BK}{^{B,(K)}}

\newcommand{\Sinf}{(\ref{eq:Sinfa})-(\ref{eq:Sinfb})-(\ref{eq:Sinfc})-(\ref{eq:Sinfd}) }
\newcommand{\Sinfe}{(\ref{eq:Sinfa})-(\ref{eq:Sinfb})-(\ref{eq:Sinfc})-(\ref{eq:Sinfd}$^{'}$) }

\title{Resource Allocation for Downlink Cellular OFDMA Systems: 
Part II---Practical Algorithms and Optimal Reuse Factor}
\author{Nassar Ksairi$^{(1)}$\footnote{$^{(1)}$Sup\'elec, Plateau de Moulon
91192 Gif-sur-Yvette Cedex, France (nassar.ksairi@supelec.fr). 
Phone: +33 1 69 85 14 54, Fax: +33 1 69 85 14 69.}, 
Pascal Bianchi$^{(2)}$\footnote{$^{(2)}$CNRS / Telecom ParisTech (ENST), 
46 rue Barrault 75634 Paris Cedex 13, France
(bianchi@telecom-paristech.fr,ciblat@telecom-paristech.fr,walid.hachem@enst.fr).
Phone: +33 1 45 81 83 60, Fax: +33 1 45 81 71 44.}, 
Philippe Ciblat$^{(2)}$, Walid Hachem$^{(2)}$}

\begin{document}
\maketitle

\begin{abstract}
In a companion paper (see Resource Allocation for Downlink Cellular
OFDMA Systems: Part~I~---~Optimal Allocation),
we characterized the optimal resource allocation in terms of power control and
subcarrier assignment, for a downlink sectorized OFDMA system impaired by
multicell interference. In our model, the network is assumed to be one
dimensional (linear) for the sake of analysis. We also assume
that a certain part of the available bandwidth is likely to be 
reused by different base stations while that the other part of the bandwidth is
shared in an orthogonal way between these base stations.
The optimal resource allocation characterized in Part I is obtained
by minimizing the total power spent by the network under the constraint
that all users' rate requirements are satisfied.
It is worth noting that when optimal resource allocation is used, 
any user receives data either in the reused bandwidth or in the protected
bandwidth, but not in both (except for at most one pivot-user in each cell).
We also proposed an algorithm that determines the optimal values of
users' resource allocation parameters.

As a matter of fact, the optimal allocation algorithm proposed in Part~I
requires a large number of operations. In the present paper, we propose a
distributed practical resource allocation algorithm with low complexity.
We study the asymptotic behavior of both this simplified resource allocation
algorithm and the optimal resource allocation algorithm of Part~I as the number
of users in each cell tends to infinity. Our analysis allows to prove that the
proposed simplified algorithm is asymptotically optimal
\emph{i.e.,} it achieves the same asymptotic transmit power as the optimal
algorithm as the number of users in each cell tends to infinity. As a byproduct
of our
analysis, we characterize the optimal value of the frequency reuse factor.
Simulations sustain our claims and show that substantial performance
improvements are obtained when the optimal value of the frequency reuse factor
is used.
\end{abstract}
\begin{keywords}
 OFDMA, Multicell Resource Allocation, Distributed Resource
Allocation, Asymptotic Analysis.
\end{keywords}
\section{Introduction}
In a companion paper~\cite{partI}, we introduced the problem of joint power
control and subcarrier assignment in the downlink of a one-dimensional
sectorized two-cell OFDMA system. Resource allocation parameters have been
characterized in such a way that {\sl i)} the total transmit power of the
network is minimum and {\sl ii)} all users' rate requirements are satisfied.
Similarly to~\cite{gau-hac-cib-1},
we investigate the case where the channel state information  at the Base Station
(BS) side is limited to some channel statistics. However, contrary
to~\cite{gau-hac-cib-1}, our model assumes that the available bandwidth is
divided into two bands: the first one is reused by different base stations
(and is thus subject to multicell interference) while the second one is shared
in an orthogonal way between the adjacent base stations (and is thus protected
from multicell interference). The number of subcarriers in each band is directly
related to the frequency reuse factor. We also assume that each user is likely
to modulate subcarriers in each of these two bands and thus we do not assume
\emph{a priori} a geographical separation of users modulating
in the two different bands. The solution to the above resource allocation
problem is given in the first part of this work. This solution turns out to be
``binary'': except for at most one pivot-user, users in each cell must be
divided into two groups, the nearest users modulating subcarriers only in the
reused band and the farthest users modulating subcarriers only in the protected
band. An algorithm that determines the optimal values of users'
resource allocation parameters is also proposed in the first part.

It is worth noting that this optimal allocation algorithm is still
computationally demanding, especially when the number of users in each cell is
large. One of the computationally costliest operations involved in the optimal
allocation is the determination of the pivot-user in each cell. In the present
paper, we propose a distributed simplified resource
allocation algorithm with low computational complexity, and we discuss its
performance as compared to the optimal resource allocation algorithm of Part~I.
This simplified algorithm assumes a pivot-distance that is fixed in advance
prior to the resource allocation process. Of course, this predefined
pivot-distance should be relevantly chosen. For that sake, we show that when the
fixed pivot-distance of the simplified algorithm is chosen according to a
certain asymptotic analysis of the optimal allocation scheme, the performance
of the simplified algorithm is close to the optimal one, provided that the
number of users in the network is large enough.
Therefore, following the approach of~\cite{gau-hac-cib-1}, we propose to
characterize the limit of the total transmit power which results from the
optimal resource allocation policy as the number of users in each cell tends to
infinity. Several existing works on resource allocation resorted to this kind of
asymptotic analysis, principally in order to get tractable formulations
of the optimization problem that can be solved analytically.
For example, the asymptotic analysis was used in~\cite{asympOFDMA1}
and~\cite{asympOFDMA2} in the context of downlink and uplink single cell OFDMA
systems respectively, as well as in~\cite{asympCDMA} in the context of
\emph{Code Division Multiple Access} (CDMA) systems with fading channels.
Another application of the asymptotic analysis can be found
in~\cite{scaling_laws}. The authors of the cited work addressed the optimization
of the sum rate performance in a multicell network. In this context, the authors
proposed a decentralized algorithm that maximizes an upper-bound on the
network sum rate. Interestingly, this upper-bound is proved to be tight in the
asymptotic regime when the number of users per cell is allowed to grow to
infinity. However, the proposed  algorithm does not guaranty fairness among the
different users. 

In this paper, we use the asymptotic analysis in order to obtain a compact
form of the (asymptotic) power transmitted by the network for the optimal
resource allocation algorithm, and we use this result to propose relevant values
of the fixed pivot-distance associated with the simplified allocation
algorithm. We prove in particular that when this fixed pivot-distance is
chosen equal to the asymptotic optimal pivot-distance, 
then the power transmitted when using the proposed simplified resource
allocation is asymptotically equivalent to the minimum power associated with the
optimal algorithm. This limiting expression no longer depends on the
particular network configuration, but on an asymptotic, or ``average'', state of
the network. More precisely, the asymptotic transmit power depends on the
average rate requirement and on the density of users in each cell. It also
depends on the value $\alpha$ of the frequency reuse factor. As a byproduct of
our asymptotic analysis, we are therefore able to determine an optimal value of
the latter reuse factor. This optimal value is defined as the value of $\alpha$
which minimizes the asymptotic power.

The rest of this paper is organized as follows. In Section~\ref{sec:sys_model}
we recall the system model as well as the joint resource allocation problem. In
Section~\ref{sec:sub_optimal_1d}, we propose a novel suboptimal distributed
resource allocation algorithm. Section~\ref{sec:asymp} is devoted to the
asymptotic analysis of the performance of this simplified allocation algorithm
as well as the performance of the optimal resource allocation scheme of Part~I
when the number of users tends to infinity.
Theorem~\ref{the:asymptotic} characterizes the
asymptotic behavior of the optimal joint allocation scheme. The results of this
theorem are used in Subsection~\ref{sec:asymp_subopt} in order to determine
relevant values of the fixed pivot-distances associated with the simplified
allocation algorithm. Provided that these relevant values are used,
Proposition~\ref{prop:subopt} states that the simplified algorithm is
asymptotically optimal. Section~\ref{sec:asym_alpha} addresses the selection of
the best frequency reuse factor. Finally, Section~\ref{sec:simus} is devoted to
the numerical illustrations of our results.
\section{System Model and Previous Results} \label{sec:sys_model}
\subsection{System Model}
\begin{figure}[h]
  \centering
  \includegraphics[width=10cm]{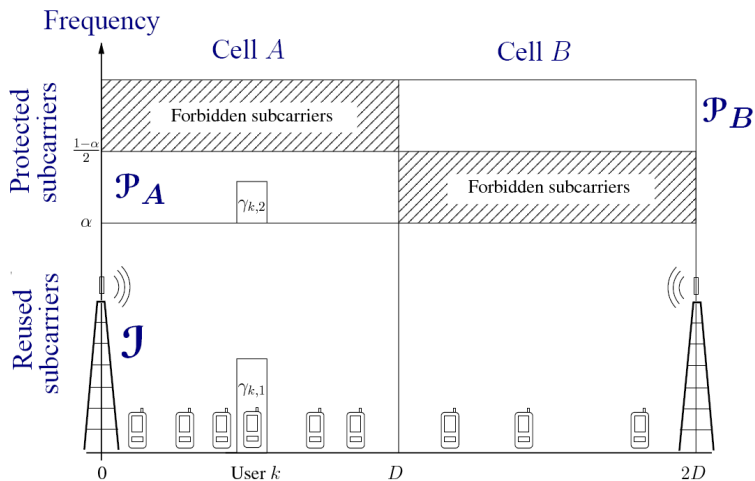}
  \caption{Two-Cell System model}
  \label{fig:model2}
\end{figure}

We consider a sectorized downlink OFDMA cellular network.
We focus on two neighboring one-dimensional (linear) cells, say Cell~$A$ and
Cell~$B$, as illustrated by Figure~\ref{fig:model2}. 
Denote by $D$ the radius of each cell. We denote by $K^A$ 
the number of users of Cell~$A$ and by $K^B$ the number of users of Cell~$B$. 
The total number of available subcarriers in the system is denoted by $N$.
For a given user $k \in {1, 2, \ldots, K^c}$ in Cell~$c$ ($c\in\{A,B\}$), 
we denote by $x_k$ the distance that separates him/her from BS~$c$,
and by ${\cal N}_k$ the set of indices corresponding to the subcarriers modulated
by $k$. ${\cal N}_k$ is a subset of $\{0, 1, \ldots,N-1 \}$.
The signal received by user~$k$ at the $n$th subcarrier 
($n \in {\cal N}_k$) and at the $m$th OFDM block is given by
\begin{equation}\label{eq:signal_model}
y_k(n,m) = H_k(n,m) s_k(n,m) + w_k(n,m), 
\end{equation} 
where $s_k(n,m)$ represents the data symbol transmitted by
BS~$c$. Process $w_k(n,m)$ is an additive noise
which encompasses the thermal noise and the possible multicell interference. 
Coefficient $H_k(n,m)$ is the frequency response of the channel at the subcarrier~$n$
and the OFDM block $m$. Random variables $H_k(n,m)$ 
are assumed Rayleigh distributed with variance $\rho_k^c=E[|H_k(n,m)|^2]$.
Channel coefficients are supposed to be perfectly known at the receiver
side, and unknown at the BS side. 
We assume that $\rho_k$ vanishes with the distance $x_k$ based on a given path loss model.
The set of available subcarriers is partitioned into three subsets: 
${\cal I}$ containing the reused subcarriers shared by the two cells; 
${\cal P}_A$ and ${\cal P}_B$ containing the protected subcarriers 
only used by users in Cell~$A$ and $B$ respectively.
The \emph{reuse factor} $\alpha$ is defined as the ratio between 
the number of reused subcarriers and the total number of subcarriers:
\begin{equation*}
\alpha=\frac{\mbox{card}(\mathcal{I})}{N}
\end{equation*}
so that $\mathcal{I}$ contains $\alpha N$ subcarriers. 
If user $k$ modulates a subcarrier $n \in \mathcal{I}$, the additive noise  
contains both thermal noise of variance $\sigma^2$ and interference. 
Therefore, the variance $\sigma_k^2$ of this noise-plus-interference process depends on $k$ 
and coincides with $\sigma_k^2=\mathbb{E}\left[|\tilde{H}_k(n,m)|^{2}\right]Q_1^B+\sigma^2$,
where $\tilde{H}_k(n,m)$ represents the channel between BS~$B$ 
and user~$k$ of Cell~$A$ at frequency $n$ and OFDM block $m$, and where
$Q_1^B=\sum_{k=1}^{K^B}\gamma\ku^BP\ku^B$ is the average power transmitted by
BS~$B$ in the interference bandwidth~$\cal I$.
The remaining $(1-\alpha)N$ subcarriers are shared by 
the two cells, Cell~$A$ and $B$~, in an orthogonal way. If user~$k$ modulates 
such a subcarrier $n \in {\cal P}_c$, the additive noise 
$w_k(n,m)$ contains only thermal noise. In other words, subcarrier $n$
does not suffer from multicell interference. Then we simply
write $\mathbb{E}[|w_k(n,m)|^2] = \sigma^2$.
The resource allocation parameters for user~$k$ are: 
$P\ku^c$ the power transmitted on each of the subcarriers of the non protected
band 
$\cal I$ allocated to him, $\gamma\ku^c$ his share of $\cal I$,
$P\kd^c$ the power transmitted on each of the subcarriers of the protected band 
$\mathcal{P}_c$ allocated to him and $\gamma\kd^c$ his share of $\mathcal{P}_{c}$.
In other words,
\begin{equation*}
\gamma\ku^c = \mbox{card}(\mathcal{I} \cap \mathcal{N}_{k})/N \qquad \gamma\kd^c = \mbox{card}(\mathcal{P}_{c} \cap \mathcal{N}_{k})/N\:.
\end{equation*}
As a consequence, $\sum_{k=1}^{K^c}\gamma\ku^c=\alpha$ and $\sum_{k=1}^{K^c}\gamma\kd^c=\frac{1-\alpha}{2}$ for
each cell $c$.
Moreover, let $g\ku$ (resp. $g\kd$) be the channel 
Gain to Noise Ratio (GNR) in band $\mathcal{I}$ (resp. $\mathcal{P}_c$), 
namely $g\ku = \rho_{k} / \sigma_k^2$ (resp. $g\kd 
= {\rho_k} / {\sigma^2}$). 
``Setting a resource allocation for cell~$c$'' means setting a value for parameters 
$\{\gamma\ku^c, \gamma\kd^c, P\ku^c, P\kd^c\}_{k=1\ldots K^c}$.
\subsection{Joint Resource Allocation for Cells~$A$ and~$B$}
\label{sec:multicell}
Assume that each user $k$ has a rate requirement of $R_k$ nats/s/Hz. 
In the first Part of this work~\cite{partI},
our aim was to jointly optimize the resource allocation for the two cells
which i) allows to satisfy all target rates $R_k$ of all users, 
and ii) minimizes the power used by the two base stations in order to achieve these rates.
For each cell $c\in\{A,B\}$, denote by $\overline{c}$ the adjacent cell ($\overline{A}=B$ and $\overline{B}=A$). 
The ergodic capacity associated with a user~$k$ in Cell~$c$ is given by 
\begin{equation}\label{eq:ergodic_capacity}
C_k=\gamma\ku^c\mathbb{E}\left[\log\left(1+g\ku(Q_{1}^{\bar c})P\ku^c
Z\right)\right]
+\gamma\kd^c\mathbb{E}\left[\log\left(1+g\kd P\kd^c Z\right)\right]\:,
\end{equation}
where $Z$ is a standard exponentially distributed random variable, and where
coefficient $g\ku(Q_1^{\bar c})$ is given by
\begin{equation}\label{eq:gnr}
g\ku(Q_{1}^{\bar c})=
\frac{\rho_k}{\mathbb{E}\left[|\tilde{H}_k(n,m)|^2\right]Q_1^{\bar c}+\sigma^2},
\end{equation}
where $\tilde{H}_k(n,m)$ represents the channel between BS~$\bar{c}$ and user~$k$ 
of Cell~$c$ at frequency~$n$ and OFDM block~$m$. 
Coefficient $g\ku(Q_{1}^{\bar c})$ represents the signal to interference plus
noise ratio in the interference band $\cal I$. We assume that users
are numbered from the nearest to the BS to the farthest. As in~\cite{partI}, the
following problem will be referred to as the joint resource allocation problem
for Cells~$A$ and~$B$: Minimize the total power spent by both base stations
$Q_T\K=\dsp \sum_{c=A,B}\sum_{k=1}^{K^c}(\gamma\ku^cP\ku^c+\gamma\kd^cP\kd^c)$
with respect to $\{\gamma\ku^c, \gamma\kd^c,P\ku^c,P\kd^c\}_{\substack{c=A,B\\
k=1\ldots K^c}}$ under the following constraint that all users' rate
requirements $R_k$ are satisfied \emph{i.e.}, for each user $k$ in any cell $c$,
$R_k\leq C_k$.
The solution to this problem has been determined in the first part of this
work~\cite{partI}. As a noticeable point, the results of~\cite{partI} indicate
the existence in each cell of a pivot-user that separates two groups of
users: the ``protected'' users and the ``non protected'' users. The following
proposition states this binary property of the solution. 
\begin{prop}[\cite{partI}]\label{prop:binary}
 Any global solution to the joint resource allocation problem
is ``binary'' \emph{i.e.,} there exists a user $L^c$ in each
Cell~$c$ such that $\gamma\kd=0$ for closest users $k<L^c$, and $\gamma\ku=0$
for farthest users $k>L^c$. 
\end{prop}
In the sequel, we denote by $d\cK$ the position of the pivot-user $L^c$ in
Cell~$c$ \emph{i.e.,} $d\cK=x_{L^c}$.
A resource allocation algorithm is also proposed in~\cite{partI}. This algorithm
turns out to have a high computational complexity and the determination of the
optimal value of the pivot-distance $d\cK$ turns out to be one of the costliest
operations involved in this algorithm. This is why we propose in the follwing
section of the present paper a suboptimal simplified allocation algorithm that
assumes a predefined pivot-distance.
\section{Practical Resource Allocation Algorithm}
\label{sec:sub_optimal_1d}
\subsection{Motivations and Main idea}
Proposition~\ref{prop:binary} provides the general form of
the optimal resource allocation, showing in particular the existence of
pivot-users $L^A$, $L^B$ in both Cells $A$, $B$, separating the users who
modulate in band $\cal I$ from the users who modulate in bands ${\cal P}^A$ and
${\cal P}^B$.
As a matter of fact, the determination of pivot-users $L^A, L^B$ is one of the
costliest operations of this optimal allocation (see~\cite{partI} for a
detailed computational complexity analysis).
Thus, it would be convenient to propose an allocation procedure for
which the pivot-position would be {\bf fixed in advance} to a constant rather
than systematically computed/optimized. We propose a simplified resource
allocation algorithm based on this idea. 
Furthermore, we prove that when the value of the fixed pivot-distances is
relevantly chosen, the proposed algorithm is asymptotically optimal as the
number of users increases. In other words, the total power spent by the network
for large $K$ when using our suboptimal algorithm does not exceed the minimum
power that would have been spent by using the optimal resource allocation. The
proposed algorithm is based on the following idea. 

Recall the definition of $d\AK$ and $d\BK$ as the respective position of the
optimal pivot-users $L^A$ and $L^B$ defined by Proposition~\ref{prop:binary}.
As the optimal pivot-positions $d\AK$ and $d\BK$ are difficult to compute
explicitly and depend on the particular rates and users' positions, we propose
to replace $d\AK$ and $d\BK$ with predefined values $d_{\textrm{subopt}}^A$ and
$d_{\textrm{subopt}}^B$ fixed before the resource allocation process. 
In our suboptimal algorithm, all users in Cell~$c$ whose distance to the BS is
less than $d_{\textrm{subopt}}^c$ modulate in the interference band $\cal I$.
Users farther than $d_{\textrm{subopt}}^c$ modulate in the protected band 
${\cal P}^c$. Of course, we still need to determine the pivot-distances
$d_{\textrm{subopt}}^A$ and $d_{\textrm{subopt}}^B$.
{\bf A procedure that permits the relevant selection of
$\boldsymbol{d_{\textrm{subopt}}^A,
d_{\textrm{subopt}}^B}$ is given in Section~\ref{sec:QT}}.
\subsection{Detailed Description}
Assume that the values of $d_{\textrm{subopt}}^A$ and $d_{\textrm{subopt}}^B$
have been fixed beforehand prior to
the resource allocation process. 
For each Cell $c$, define by ${\cal K}^c_I$ the subset of $\{1,\dots K^c\}$
corresponding to the users whose distance to BS~$c$ is less than
$d_{\textrm{subopt}}^c$. Define
by ${\cal K}^c_P$ the set of users whose distance to BS~$c$ is larger than
$d_{\textrm{subopt}}^c$. 
\subsubsection{Resource allocation for protected users}
\label{sec:alloc_for_noninter}
Focus for instance on Cell~$A$. For each $k\in {\cal K}^A_P$, we arbitrarily
set $\gamma\ku^A=P\ku^A=0$ \emph{i.e.}, user $k$ is forced to modulate in the
protected band ${\cal P}^A$ only. For such users, the remaining resource
allocation parameters $\gamma\kd^A, P\kd^A$ are obtained by solving the
following classical single cell problem w.r.t.
$(\gamma\kd^A, P\kd^A)_{k\in {\cal K}^A_P}$: 
\begin{center}
  ``Minimize the transmitted power $\sum_{k\in {\cal K}^A_P} \gamma\kd^A P\kd^A$
under rate constraint
  $R_k<C_k$ for each $k\in {\cal K}^A_P$''.
\end{center}
The above problem is a simple particular case of the single cell problem 
addressed in~\cite{partI}. Define the functions
$f(x) = \frac{\EE{\log(1+xZ)}}{\EE{\frac{Z}{1+xZ}}}-x$ and
$C(x)=\mathbb{E}[\log(1+f^{-1}(x)Z)]$ on ${\mathbb{R}}_+$. 
The solution is given by
\begin{align*}
  &  P\kd^A=g\kd^{-1}f^{-1}(g\kd \tilde \beta_2) \\
  &  \gamma\kd^A=\frac{R_k}{\EE{\log\left(1+g\kd P\kd^A Z\right)}}\:,
\end{align*}
where parameter $\tilde \beta_2$ is obtained by writing that constraint 
$\sum_k \gamma\kd^A = \frac{1-\alpha}{2}$ holds or equivalently, $\tilde
\beta_2$ is the unique solution to:
$$
\sum_{k\in {\cal K}^A_P} \frac{R_k}{C(g\kd \tilde \beta_2)} =
\frac{1-\alpha}{2}\:.
$$
We proceed similarly for Cell~$B$.

\subsubsection{Resource allocation for interfering users}
\label{sec:alloc_for_inter}

We now focus on users $k\in {\cal K}^c_I$ for each cell $c=A,B$. For such
users, we arbitrarily set $\gamma\kd^c=P\kd^c=0$ \emph{i.e.}, users in ${\cal
K}^c_I$ are forced to modulate in the interference band ${\cal I}$ only, for
each cell $c$. The remaining resource allocation parameters $\gamma\ku^A,
P\ku^A, \gamma\ku^B, P\ku^B$
are obtained by solving the following simplified multicell problem. 
\begin{problem}
\label{prob:opt_multi}
{\bf [Multicell]} Minimize $\dsp \sum_{c=A,B} \sum_{k\in {\cal K}_I^c}
\gamma\ku^c P\ku^c$ w.r.t. 
$(\gamma\ku^A, P\ku^A, \gamma\ku^B, P\ku^B)_k$ under the following constraints
for each cell $c\in\{A,B\}$:
\begin{align*}
\mathbf{C1:}\: \forall c, \; \forall k\in{\cal K}_I^c, R_k\leq C_k & &
\mathbf{C2:}\: \forall c, \; \sum_{k\in{\cal K}_I^c} \gamma\ku^c= \alpha & & 
\mathbf{C3:}\: \gamma\ku^c\geq0\:.
\end{align*}
\end{problem}
Clearly, the above Problem can be interpreted as a particular case 
of the initial resource allocation (Problem~2 in~\cite{partI}) addressed in
Section~\ref{sec:multicell} of the present paper. The main difference is that
the initial multicell
problem jointly involves the resource allocation parameters in three bands $\cal
I$, ${\cal P}^A$ and ${\cal P}^B$ whereas the present problem only optimizes the
resource allocation parameters corresponding to band $\cal I$, while arbitrarily
setting the others to zero. Therefore, the results of
Part~I~\cite{partI}, Theorem~2 of~\cite{partI} in particular, can
directly be used to determine the global solution to
Problem~\ref{prob:opt_multi}.
\begin{remark}[{\bf Feasibility}]
Recall that the initial joint resource allocation Problem (Problem~2
in~\cite{partI}) described in
Section~\ref{sec:multicell} in the present paper was always feasible.
Intuitively, this was due to
the fact that any user was likely to modulate in the protected band if needed,
so that any rate requirement $R_k$ was likely to be satisfied by simply
increasing the power in the protected band. In the present case, the protected
band is by definition forbidden to users in ${\cal K}_I^c$. Theoretically
speaking, Problem~\ref{prob:opt_multi} might not be feasible due to  multicell
interference. Fortunately, we will see this case does not happen, at least for
a sufficiently large number of users, if the values of the pivot-distances
$d_{\textrm{subopt}}^A$ and $d_{\textrm{subopt}}^B$ are well chosen. This point
will be discussed in more detail in Section~\ref{sec:subopt_feasibility}.
\label{rem:feasibility}
\end{remark}
Define $Q_1^c = \sum_{k\in{\cal K}_I^c} \gamma\ku^cP\ku^c$ as the
average power transmitted by BS~$c$ in the interference bandwidth~$\cal I$.
By straightforward application of Theorem~\ref{the:multi}, we obtain that for
each Cell~$c$ and for each user $k\in {\cal K}_I^c$, 
\begin{align}
  &  P\ku^c=g\ku^{-1}(Q_1^{\bar c}) f^{-1}(g\ku(Q_1^{\bar c}) \tilde \beta_1^c) 
\label{eq:resSimpa}\\
  &  \gamma\ku^c=\frac{R_k}{\EE{\log\left(1+g\ku(Q_1^{\bar c}) P\ku^c
Z\right)}}\:,
\label{eq:resSimpb}
\end{align}
where for each $c=A,B$ and for a fixed value of $Q_1^{\bar c}$, parameters 
$(\tilde \beta_1^c, Q_1^c)$ are the unique solution to the following system of
equations:
\begin{align}
  & \sum_{k\in{\cal K}_I^c} \frac{R_k}{C(g\ku(Q_1^{\bar c}) \tilde \beta_1^c)} =
\alpha \label{eq:simpa}\\
  & Q_1^c = \sum_{k\in{\cal K}_I^c} R_k \frac{g\ku^{-1}(Q_1^{\bar c})
f^{-1}(g\ku(Q_1^{\bar c}) \tilde \beta_1^c)}{C(g\ku(Q_1^{\bar c}) \tilde
\beta_1^c)}\:. 
\label{eq:simpb}
\end{align}
Note that the first equation is nothing else that the constraint 
$\bf C2$: $\sum_k \gamma\ku^c = \alpha$. The second equation is nothing else
than the definition $Q_1^c = \sum_{k\in{\cal K}_I^c} \gamma\ku^cP\ku^c$. We now
prove that the system of four equations~(\ref{eq:simpa})-(\ref{eq:simpb}) for
$c=A,B$ admits a unique solution $\tilde \beta_1^A, Q_1^A,\tilde \beta_1^B,
Q_1^B$ and we provide a simple algorithm allowing to determine this solution.

Focus on a given Cell~$c$ and consider any fixed value $Q_1^{\bar c}$. 
Denote by $\tilde I^c(Q_1^{\bar c})$ the rhs of equation~\eqref{eq:simpb} where
$\tilde \beta_1^c$ is defined as the unique solution to~\eqref{eq:simpa}.
Since \eqref{eq:simpb} should be satisifed for both $c=A$ and $c=B$, the
following two equations hold
\begin{equation*}
Q_1^A=\tilde I^A(Q_1^B),\quad Q_1^B=\tilde I^B(Q_1^A)\:.
\end{equation*}
The couple $(Q_1^A, Q_1^B)$ is therefore clearly a fixed point of the
vector-valued function $\tilde{\bf I}(Q_1^A, Q_1^B) = (\tilde I^A(Q_1^B), \tilde
I^B(Q_1^A))$.
 \begin{equation}
   \label{eq:fixedtilde}
 (Q_1^A, Q_1^B) = \tilde{\bf I}(Q_1^A, Q_1^B)\:.   
 \end{equation}
As a matter of fact, it can be shown that such a fixed point of 
$\tilde {\bf I}$ is unique. This claim can be proved using the approach
previously proposed by~\cite{yates}. 
\begin{lemma}
\label{lem:yates}  
Function $\tilde{\bf I}$ is such that the following properties hold.
  \begin{enumerate}
  \item Positivity: $\tilde{\mathbf{I}}(Q^A,Q^B)>0$.
  \item Monotonicity: If $Q^A\geq {Q^A}', Q^B\geq{Q^B}'$, then 
  $\tilde{\mathbf{I}}(Q^A,Q^B)\geq\tilde{\mathbf{I}}({Q^A}',{Q^B}')$.
  \item Scalability: for all $t>1$,
$t\tilde{\mathbf{I}}(Q^A,Q^B)>\tilde{\mathbf{I}}(tQ^A,tQ^B)$.
  \end{enumerate}
\end{lemma}
The proof of Lemma~\ref{lem:yates} uses arguments which are very similar to
the proof of Theorem~1 in~\cite{papandriopoulos}. It is thus omitted from this
paper and provided in \cite{rapport}.
Function $\tilde{\mathbf{I}}$ is then a \emph{standard interference function},
using the terminology of~\cite{yates}. Therefore, as stated in~\cite{yates},
such a function $\tilde{\bf I}$ admits at most one fixed
point. On the other hand, the existence of a fixed point is ensured by the
feasibility of Problem~\ref{prob:opt_multi} and by the fact that
(\ref{eq:fixedtilde}) holds for any global solution. In other words, if
Problem~\ref{prob:opt_multi} is feasible, then function $\tilde{\bf I}$ does
admit a fixed point and this fixed point is unique.
Putting all pieces together, there exists a unique solution
to~(\ref{eq:fixedtilde}), which can be
obtained thanks to a simple fixed point algorithm. In practice, resource
allocation in band $\cal I$ can be achieved by the following procedure.

\noindent {\bf Ping-pong algorithm for interfering users}
\begin{enumerate}
\item Initialization: $Q_1^B=0$.
\item Cell A: Given the current value of the power $Q_1^B$ transmitted by base
      station~B in the interference bandwidth, compute $\tilde\beta_1^A, Q_1^A$
      as the unique solution to~(\ref{eq:simpa})-(\ref{eq:simpb}) with $c=A$. 
\item Cell B: Given the current value of $Q_1^A$, compute $\tilde\beta_1^B,
Q_1^B$ by~(\ref{eq:simpa})-(\ref{eq:simpb}).
\item Go back to step 2 until convergence.
\item Define resource allocation parameters
by~(\ref{eq:resSimpa})-(\ref{eq:resSimpb}).
\end{enumerate}

\noindent {\bf Comments}
\begin{enumerate}
\item {\bf Convergence of the ping-pong algorithm.} 
We stated earlier that Problem~\ref{prob:opt_multi} is either feasible or
infeasible, depending on the value of
$(d_{\textrm{subopt}}^A,d_{\textrm{subopt}}^B)$. 
If the latter problem is feasible, then function $\tilde{\bf I}$ will heve a
unique fixed point due to Lemma~\ref{lem:yates} and the ping-pong algorithm will
converge to this fixed point. If Problem~\ref{prob:opt_multi} is infeasible,
then function $\tilde{\bf I}$ will have no fixed points and the the ping-pong
algorithm will diverge. One of the main purposes of Section~\ref{sec:QT} is to
provide relevant values of $(d_{\textrm{subopt}}^A,d_{\textrm{subopt}}^B)$ such
that convergence of the ping-pong algorithm holds for sufficiently large number
$K$ of users.
\item Note that the only information needed by Base Station~$c$ about
Cell~$\overline{c}$ is the current value of the power $Q_1^{\overline{c}}$
transmitted by Base Station~$\overline{c}$ in the interference band $\cal I$.
This value can \emph{i)} either be measured by Base Station~$c$ at each
iteration of the ping-pong algorithm, or \emph{ii)} it can be communicated to
it by Base Station~$\overline{c}$ over a dedicated link. In the first case, no
message passing is required, and in the second case only few information is
exchanged between the base stations. The ping-pong algorithm can thus
be implemented in a distributed fashion.
\end{enumerate}

\subsection{Complexity Analysis}

We showed earlier that allocation for protected users can be reduced to the
determination in each cell of the value of $\tilde{\beta}_2^c$, which is the
unique solution to the equation $\sum_{k\in {\cal K}^A_P} \frac{R_k}{C(g\kd
\tilde \beta_2^c)} = \frac{1-\alpha}{2}$. We argued in~\cite{partI} that solving
this kind of equations requires a computational complexity proportional to the
number of terms in the lhs of the equation, which is itself of order $O(K)$.
Using similar arguments, we can show that each iteration of the ping-pong
algorithm for non protected users can be performed with a complexity of order
$O(K)$. Let $J$ designate the number of iterations needed till convergence. The
overall computational complexity of the ping-pong algorithm, and hence of the
simplified resource allocation scheme as well, is thus of the order of $O(J K)$.
Our simulations showed that the ping-pong algorithm converges relatively quickly
in most of the cases. Indeed, no more than $J=15$ iterations were needed in
almost all the simulations settings to reach convergence within a very
reasonable accuracy.
The complexity of the simplified algorithm is to be compared with the
computational complexity of the
optimal algorithm which was shown in~\cite{partI} to be of the order of 
$O(M K \log_2 K)$, where $M$ is the number of points inside a certain 2D
search grid.
\section{Asymptotic Optimality of the Simplified Resource Allocation Scheme}
\label{sec:asymp}
The aim of this section is to evaluate the performance of the proposed
simplified algorithm. The relevant performance metric in the context of this
paper is the total power that must be transmitted by the base stations.
Since the simplified algorithm assumes predefined pivot-distances 
$(d_{\textrm{subopt}}^A,d_{\textrm{subopt}}^B)$ fixed prior to
the resource allocation process, the performance of the proposed algorithm
depends on the choice of these fixed pivot-distances. One must therefore
determine what relevant value should be selected for
$(d_{\textrm{subopt}}^A,d_{\textrm{subopt}}^B)$. A possible method is addressed
in this section and consists in studying the case where the number of users
tends to infinity.
\subsection{Main Tools: Asymptotic analysis}
\label{sec:motivation}
We study first the performance of the {\bf optimal} allocation
algorithm proposed in Part~I~\cite{partI} when the number of users in each cell
tends to infinity. From the results of this asymptotic study, we conclude the
asymptotic behaviour of the optimal pivot-distances $\left(d\AK,d\BK\right)$.
It turns out that when the number $K$ of users increases, the optimal
pivot-distances as well as the total transmitted power no longer depend
on the particular cell configuration, but on an asymptotic state
of the network, such as the average rate requirement and the density of users in
each cell.
Thanks to this result, we can now choose the fixed
pivot-distances associated with the simplified algorithm to be equal to the
asymptotic pivot-distances. In this case, one can show that the performance gap
between the simplified and the optimal allocation schemes vanishes for high
numbers of
users. We introduce now the mathematical assumptions and tools that we use for
defining the asymptotic regime.
\subsubsection{Notations and Basic Assumptions}
In the sequel, we denote by $B$ the total bandwidth of the system in Hz. 
We consider the asymptotic regime where
the number of users in each cell tends to infinity. 
We denote by $r_k=BR_k$ the data rate requirement of user $k$ in nats/s, and we
recall that $R_k$ is the data rate requirement of user~$k$ in nats/s/Hz. Notice
that the total rate $\sum_{k=1}^{K^c} r_k$ which should be delivered by BS~$c$
tends to infinity as well. Thus, we need to let the bandwidth $B$ grow to
infinity in order to satisfy the growing data rate requirement. Recalling that
$K=K^A+K^B$ 
denotes the total number of users in both cells, 
the asymptotic regime will be characterized by $K \to \infty,B \to \infty$ and
$K/B \to t$ where $t$ is a positive real number. We assume on the other hand
that $K^c/K$ ($c\in\{A,B\}$) tends to some positive constant as $K$ tends to
infinity. Without restrictions, this constant is assumed in the sequel to be
equal to 1/2 {\sl i.e.}, the number of users becomes equivalent in each cell. In
order to simplify the proofs of our results, we assume without restriction that
for each $k$, the rate requirement $r_k$ is upper-bounded by a certain constant
$r_{\max}$, $r_k\leq r_{\max}$, where $r_{\max}$ can be chosen as large as
needed, and that users of each cell are located in the interval $[\epsilon,D]$
where $\epsilon>0$ can be chosen as small as needed. Recall that $x_k$ denotes
the position of each user $k$ {\sl i.e.,} the distance between the user and the
BS. The variance of the channel gain of user $k$ will be written as
$\rho_k=\rho(x_{k})$ where $\rho(x)$ models the path loss. Typically, function
$\rho(x)$ has the form $\rho(x)=\lambda x^{-s}$ where $\lambda$ is a certain
gain and where $s$ is the path-loss coefficient, $s \geq 2$. In the sequel, we
denote by $g_2(x) = \frac{\rho(x)}{\sigma^2}$ the received gain to noise ratio
in the protected bandwidth, for a user at position $x$. This way, $g_2(x_k) =
g\kd$. Similarly, we define for each user $k$ in cell $A$, $g_1(x_k,Q_1^B) =
g\ku(Q_1^B)$. More generally, $g_1(x,{\cal Q})$ denotes the
gain-to-interference-plus-noise ratio  in the interference bandwidth at position
$x$ when the interfering cell is transmitting with power $\cal Q$ in band $\cal
I$. Functions $g_1(x,.)$ and $g_2(x)$ are assumed to be continuous functions of
$x$. It is worth noting that for each $x$, $g_2(x)=g_1(x,0)$. Finally, recall
that coefficient $\gamma\ku^c$ (resp. $\gamma\kd^c$) is defined as the ratio
between the part of the interference bandwidth $\cal I$ (resp. protected
bandwidth ${\cal P}_c$) and the total bandwidth. Thus, $\gamma\ku^c$ and
$\gamma\kd^c$ tend to zero as the total bandwidth $B$ tends to infinity for each
$k$. 
\subsubsection{Statistical Tools and Main Ideas of the Asymptotic Study}
\label{sec:tools}

Theorem~2 of Part~I~\cite{partI} reduces the
determination of the whole set of resource allocation
parameters in both cells to the determination of ten unknown
parameters $\{Q_1^c,\beta_i^c,L^c,\xi^c\}_{c=A,B,\: i=1,2}$.
Parameter $Q_1^c$ in particular represents the power transmitted by Cell~$c$ in
the non protected band $\cal I$.
Consider now one of the two Cells $c\in\{A,B\}$, and denote by $\overline{c}$ 
the second (adjacent) cell. In the sequel, we use the notation $Q_1\cK$ (resp.
$Q_2\cK$) instead of $Q_1^c$ (resp. $Q_2^c$) to designate the power transmitted
by BS~$c$ in the non protected band $\cal I$
(resp. the protected band $\mathcal{P}_c$) when the optimal solution
characterized by Proposition~\ref{prop:binary} is used.
\begin{eqnarray} 
Q_1\cK&=&\sum_{k=1}^{L^c}\gamma\ku^c P\ku^c \label{eq:opt_q1}\\ 
Q_2\cK&=&\sum_{k=L^c}^{K^c}\gamma\kd^c P\kd^c\:.\label{eq:opt_q2}
\end{eqnarray}
The new notation $Q_1\cK,Q_2\cK$ is used to indicate the dependency of 
the results on the number of users $K$.
For the same reason, parameters
$L^c,\beta_1^c,\beta_2^c,\xi^c$ will be denoted in the sequel by
$L\cK$, $\beta_1\cK$, $\beta_2\cK$, $\xi\cK$ respectively. Our goal now is to
characterize the behavior of the resource allocation strategy as $K,B\to\infty$
and, in particular, the behavior of powers $Q_1\cK$, $Q_2\cK$. By
straightforward application of Theorem~2 of Part~I,
$Q_1\cK=\sum_{k=1}^{L^c}\gamma\ku^c P\ku^c$ can be written as 
\begin{equation}
Q_1^{c,(K)}= \sum_{k<L\cK} R_k 
{\cal F}(x_k,\beta_1\cK,Q_1^{\bar{c},(K)},\xi\cK)
+W_{L^{c,(K)},1}^c\label{eq:finite_Q1}\:,
\end{equation}
where $W_{L\cK,1}^c = \gamma_{L\cK,1}^cP_{L\cK,1}^c$ denotes the 
power transmitted to the pivot-user $L\cK$ in the interference band $\cal I$,
and where function $\cal F$ is defined by
\begin{equation}\label{eq:calF}
{\cal F}(x,\beta,{\cal Q},\xi)=\frac{f^{-1}\left(\frac{g_1(x,{\cal
Q})}{1+\xi}\beta\right)}{g_1(x,{\cal Q})
C\left(\frac{g_1(x,{\cal Q})}{1+\xi}\beta\right)}
\end{equation}
for each $x,\beta,{\cal Q}$. The first term in the rhs of~(\ref{eq:finite_Q1}) 
represents the total power allocated to all users $k<L^{c,(K)}$. It is quite
intuitive that the power allocated to one user $W_{L\cK,1}^c$ 
is negligible when compared to the power allocated to all users $k<L\cK$. 
Indeed, it will be shown in Appendix~\ref{app:asym_proof} that the first term
of~\eqref{eq:finite_Q1} is bounded as $K\to\infty$ wheras $W_{L\cK,1}^c$ tends
to zero. In the sequel, we use notation $W_{L\cK,1}^c = o_K(1)$, where $o_K(1)$ 
stands for any term which converges to zero as $K\to\infty$.
In order to study the limit of this expression as $K$ tends to infinity, we
introduce for each one of the two cells the following measure $\nu\cK$ defined
on the Borel sets of $\mathbb{R}_+\times\mathbb{R}_+$ as follows
\begin{equation}\label{eq:K_measure}
\nu\cK(I,J) =\frac{1}{K^c}\sum_{k=1}^{K^c}\delta_{r_k,x_k}(I,J)
\end{equation}
where $I$ and $J$ are any intervals of $\mathbb{R}_+$ and where 
$\delta_{r_k,x_k}$ is the Dirac measure at point $(r_k,x_k)$. In order to have
more insights on the meaning of this tool, it is useful to remark that
$\nu\cK(I,J)$ is equal to
$$
\nu\cK(I,J)=\frac{\textrm{number of users located in }J
\textrm{ and requiring a rate (in nats/s) in interval }I}{\textrm{total number
of users}}.
$$
Thus, measure $\nu\cK$ can be interpreted as the distribution of 
the set of couples $(r_k,x_k)$ of Cell~$c$.
The introduction of the above measure simplifies considerably 
the asymptotic study of the transmit power.
Indeed, replacing $R_k$ (in nats/s/Hz) by $\frac{r_k\mbox{ (nats/s) }}{B}$ 
in equation~(\ref{eq:finite_Q1}), we obtain 
\begin{eqnarray}
  Q_1\cK &=& \frac{1}{B}\sum_{k < L\cK}r_k{\cal
F}(x_k,\beta_1\cK,Q_1^{\bar{c},(K)},\xi\cK) + o_K(1)\nonumber\\
  &=& \frac{K^c}{B} \int\!\!\!\! \int_{\Delta_1\cK} r{\cal
F}(x,\beta_1\cK,Q_1^{\bar{c},(K)},\xi\cK)d\nu\cK(r,x)+ o_K(1)\:,
  \label{eq:fonctionMesure}
\end{eqnarray} 
where integration is considered with respect to the set 
$\Delta_1\cK=[0,r_{\max}]\times [\epsilon, d\cK]$,
where $d\cK=x_{L\cK}$ is the position of pivot-user $L\cK$ 
and where $\epsilon$ can be chosen, as stated earlier in this section, as small
as needed. It is quite intuitive that the 
asymptotic power $\lim_{K\to\infty} Q_1\cK$ can be obtained
from~\eqref{eq:fonctionMesure} by replacing 
$\frac{K^c}{B}=\frac{K}{B}\times \frac{K^c}{K}$ by $t\times \frac{1}{2}$ 
and the distribution $\nu\cK$ by the \emph{asymptotic distribution} $\nu^c$ of
couples $(r_k,x_k)$ as $K$ tends to infinity.
The existence and the definition of this asymptotic distribution is provided by
the following assumption.
\begin{assumption}
\label{ass:measure_convergence}
As $K$ tends to infinity, measure $\nu\cK$ converges weakly to a measure
$\nu^c$.
\end{assumption}
We refer to \cite{bil-1} for the materials on the convergence of measures.
In order to have some insight on the behavior of
equation~\eqref{eq:fonctionMesure} in the asymptotic regime, imagine for
the sake of simplicity  that sequences 
$d\AK,d\BK,Q_1\AK$, $Q_1\BK,\beta_1\AK,\beta_1\BK,\xi\AK,\xi\BK$
are convergent and that they converge respectively to 
$d^A,d^B,Q_1^A$, $Q_1^B,\beta_1^A,\beta_1^B,\xi^A,\xi^B$.
This assumption is of course arbitrary for the moment, but it allows to better
understand the main ideas of our asymptotic analysis. More rigorous
considerations on the convergence of these sequences will be
discussed later on. Ignoring at first such technical issues, it is
intuitive from equation~\eqref{eq:fonctionMesure} that $Q_1\cK$ converges
to a constant $Q_1^c$ defined by
\begin{equation}\label{eq:Q1d_nu}
Q_1^c=\frac{t}{2} \int\!\!\!\! \int_{\Delta_1^c} r{\cal
F}(x,\beta_1^c,Q_1^{\bar{c}},\xi^c)d\nu^c(r,x) \:,
\end{equation}
where $\Delta_1^c=[0,r_{\max}]\times [\epsilon, d^c]$. In other words, 
we manage to express the limit of the power $Q_1\cK$ transmitted by station $c$
in the interference band as a function of the asymptotic cell configuration. 
In order to further simplify the above expression, it is also realistic to
assume that measure $\nu^c$ is the measure product of a limit rate distribution
times a limit location distribution. Assumption~\ref{ass:measure_product} below
is motivated by the observation that in practice, the rate requirement $r_k$ of
a given user is usually not related to the position $x_k$ of the user in each
cell.
\begin{assumption}\label{ass:measure_product}
Measure $\nu^c$ is such that $d\nu^c(r,x)=d\zeta^c(r)\times d\lambda^c(x)$ 
where $\zeta^c$ is the limit distribution of rates and $\lambda^c$ is the limit
distribution of the users' locations. Here $\times$ denotes the product of
measures.
\end{assumption}
Measures $\zeta$ and $\lambda$ respectively correspond to the distributions 
of the rates and the positions of the users within one cell. For instance, the
value $\bar{r}^c=\frac{t}{2}\int_{0}^{r_{\max}}r\: d\zeta^c(r)$ represents the
average rate requirement per channel use in Cell~$c$. We furthermore assume that
measures $\lambda^A$ and $\lambda^B$ are absolutely
continuous with respect to the Lebesgue measure on $[\epsilon,D]$.
Using Assumption~\ref{ass:measure_product}, equation~\eqref{eq:Q1d_nu} becomes
\begin{equation}\label{eq:Q1d}
Q_1^c = {\bar r}^c\int_\epsilon^{d^c}{\cal
F}(x,\beta_1^c,Q_1^{\bar{c}},\xi^c)\:d\lambda^c(x).
\end{equation}
Of course, a similar result can be obtained for $Q_2\cK$ \emph{i.e.},
the power transmitted by base station $c$ in the protected band ${\cal P}^c$.
To that end, we simply note that function $g_2(x)$ satisfies
$g_2(x)=g_1(x,0)$. Using similar tools, the expression of $Q_2\cK$ given
by~\eqref{eq:multisup}
converges as $K\to\infty$ toward
\begin{equation}\label{eq:Q2d}
Q_2^c = {\bar r}^c\int_{d^c}^D{\cal F}(x,\beta_2^c,0,0)\:d\lambda^c(x).
\end{equation}
Equations~\eqref{eq:Q1d} and~\eqref{eq:Q2d} respectively provide the limits of
$Q_1\cK$ 
and $Q_2\cK$ as a function of some parameters $d^c,\beta_1^c,\beta_2^c$ and
$Q_1^{\bar c}$
(assumed for the moment to be the limits of $d\cK,\beta_1\cK,\beta_2\cK$ and
$Q_1^{\bar c,(K)}$
as long as such limits exist).
These unknown parameters still need to be characterized.
Therefore, we must determine a system of equations which is satisfied
by these parameters. This task is done by Theorem~\ref{the:asymptotic} given
below.
\subsection{Asymptotic Performance of the Optimal Resource Allocation}
\label{sec:asymp_opt}

Define the following function ${\cal G}(x,\beta,{\cal
Q},\xi)=\frac{1}{C\left(\frac{g_1(x,{\cal
Q})}{1+\xi}\beta\right)}$ for each $x,\beta,{\cal Q},\xi$.
The proof of the following result is provided in Appendix~\ref{app:asym_proof}.
\begin{theo}
\label{the:asymptotic}
Assume that $K=K^A+K^B\to \infty$ in such a way that $K/B\to t>0$ and $K^A/K\to
1/2$. Assume that the optimal solution for the joint
resource allocation problem (Problem~2 in~\cite{partI}) is used for each $K$.
The total power spent by the network
$Q_T\K=\sum_{c=A,B}\sum_{k=1}^{K^c}(\gamma\ku^c P\ku^c+\gamma\kd^c P\kd^c)$ 
converges to a constant $Q_T$. The limit $Q_T$ has the following form:
\begin{eqnarray}
  Q_T &=& \sum_{c=A,B} {\bar r}^c\left(\int_\epsilon^{d^c}{\cal
F}(x,\beta_1^c,Q_1^{\bar{c}},\xi^c)\:d\lambda^c(x) + 
  \int_{d^c}^D{\cal F}(x,\beta_2^c,0,0)\:d\lambda^c(x)\right)\:,
\label{eq:limitQT}
\end{eqnarray}
where for each $c=A,B$, the following system of equations 
in variables $d^c,\beta_1^c,\beta_2^c,\xi^c$ is satisfied:
\begin{align}
 & {\bar r}^c\int_\epsilon^{d^c}{\cal
G}(x,\beta_1^c,Q_1^{\bar{c}},\xi^c)\:d\lambda^c(x)=\alpha
\label{eq:Sinfa}\\
 & {\bar r}^c\int_{d^c}^D{\cal
G}(x,\beta_2^c,0,0)\:d\lambda^c(x)=\frac{1-\alpha}{2} 
\label{eq:Sinfb}\\
 & \frac{g_1(d^c,Q_1^{\bar c})}{1+\xi^c} F\left(\frac{g_1(d^c,Q_1^{\bar
c})}{1+\xi^c} \beta_1^c\right)=
 g_2(d^c) F(g_2(d^c)\beta_2^c)
\label{eq:Sinfc}\\
 & {\bar r}^c\int_\epsilon^{d^c}{\cal
F}(x,\beta_1^c,Q_1^{\bar{c}},\xi^c)\:d\lambda^c(x) = Q_1^c\:.
\label{eq:Sinfd}
\end{align}
Moreover, for each $c=A,B$ and for any arbitrary fixed value
$(\tilde{Q}_1^A,\tilde{Q}_1^B)$, 
the system of equations \Sinf admits at most one solution
$(d^c,\beta_1^c,\beta_2^c,\xi^c)$.
\end{theo}
As a consequence, when optimal multicell resource allocation is used, 
the total power spent by the network converges to a constant which can be
evaluated through the
results of Theorem~\ref{the:asymptotic}. This result allows to evaluate the
asymptotic power spent by the network as a function of the reuse factor
$\alpha$, the average rate requirement $\bar r$
and the asymptotic distribution of users in each cell $\lambda$.

Now that the asymptotic performance of the optimal allocation scheme has been
studied, the value of the fixed pivot-distances
$d_{\textrm{subopt}}^A,d_{\textrm{subopt}}^B$ associated with the simplified
allocation algorithm can be relevantly chosen to be equal in each Cell~$c$ to
the asymptotic pivot distance $d^c$ defined by Theorem~\ref{the:asymptotic}. 
\subsection{Determination of the fixed pivot-distances 
$d_{\textrm{subopt}}^A,d_{\textrm{subopt}}^B$ for the simplified allocation
scheme}
\label{sec:QT}
We stated earlier in Section~\ref{sec:sub_optimal_1d} that the suboptimal
algorithm replaces the optimal value $d\cK$ of the pivot-distance in
each Cell~$c$ with a fixed value $d_{\textrm{subopt}}^c$.
Intuitively, if $d_{\textrm{subopt}}^A$ and $d_{\textrm{subopt}}^B$ are chosen
such that $d\AK\simeq d_{\textrm{subopt}}^A$ and
$d\BK\simeq d_{\textrm{subopt}}^B$ for large $K$, the performance of our
algorithm shall be close to the optimal one as $K$ increases.
Therefore, we must determine an asymptotically optimal pair of pivot-distances
$(d^A,d^B)$. To that end we propose the following procedure.

Note first by referring to Theorem~\ref{the:asymptotic} 
that the value of $d^A, d^B$ can be easily determined once the
relevant values of $Q_1^A$ and $Q_1^B$ have been determined. 
The remaining task is thus the determination of the value of $(Q_1^A,Q_1^B)$.
To that end, we propose to perform an exhaustive search on~$(Q_1^A,Q_1^B)$.

\noindent {\sl i)} For each point $(\tilde Q_1^A,\tilde Q_1^B)$ on a certain
2D search grid, solve the system~\Sinf introduced by
Theorem~\ref{the:asymptotic} for both
$c=A,B$. Theorem~\ref{the:asymptotic} states that this system admits at most one
solution for any arbitrary fixed value $(\tilde Q_1^A,\tilde Q_1^B)$.
If the investigated point $(\tilde Q_1^A,\tilde Q_1^B)$ of the grid is such that
the system \Sinf does admit a solution, we can obtain
this solution denoted by $d^c(\tilde Q_1^A,\tilde Q_1^B)$, $\beta_1^c(\tilde
Q_1^A,\tilde Q_1^B)$, $\beta_2^c(\tilde Q_1^A,\tilde Q_1^B)$, $\xi^c(\tilde
Q_1^A,\tilde Q_1^B)$ thanks to a simple procedure inspired by the {\sl
single-cell} procedure proposed in Part~I~\cite{partI} for finite
number of users:
\begin{itemize}
\item Solve the system~\Sinfe formed by replacing the equality in
equation~\eqref{eq:Sinfd} of system~\Sinf by the following inequality
\begin{equation}
{\bar r}^c\int_\epsilon^{d^c}{\cal
F}(x,\beta_1^c,Q_1^{\bar{c}},\xi^c)d\lambda^c(x) \leq \tilde Q_1^c\:.
\tag{\ref{eq:Sinfd}$^{'}$}
\end{equation}
The existence and the uniqueness of the solution to this new system for 
an arbitrary $(\tilde Q_1^A,\tilde Q_1^B)\in\mathbb{R}_+^2$ can be proved by
extending, to the case of infinite number of users, Proposition~1
which was provided in~\cite{partI} for the case of finite number of users.

\item If the resulting power ${\bar r}^c\int_\epsilon^{d^c}{\cal
F}(x,\beta_1^c,Q_1^{\bar{c}},\xi^c)d\lambda^c(x)$ 
transmitted in the interference band ${\cal P}_c$ is equal to $\tilde Q_1^c$, 
then the resulting value of $d^c(\tilde Q_1^A,\tilde Q_1^B)$ coincides with the
unique solution to system~\Sinf. Once again, this claim can be proved by
extending Proposition~1 of~\cite{partI} to the case of infinite number of users.

\item If the power ${\bar r}^c\int_\epsilon^{d^c}{\cal
F}(x,\beta_1^c,Q_1^{\bar{c}},\xi^c)d\lambda^c(x)$ 
is less than $\tilde Q_1^c$, 
then $d^c(\tilde Q_1^A,\tilde Q_1^B)$ is clearly not a solution to
system~\Sinf, as equality~\eqref{eq:Sinfd} does not hold. 
In this case, it can be easily shown that system~\Sinf has no solution.
The point $(\tilde Q_1^A,\tilde Q_1^B)$ is thus eliminated.
\end{itemize}

\noindent {\sl ii)} Compute the total power
$$
Q_T(\tilde Q_1^A,\tilde Q_1^B) = \sum_{c=A,B}\sum_k \gamma\ku^c
P\ku^c+\gamma\kd^c P\kd^c
$$
that would be transmitted if the values of $Q_1^A$ and $Q_1^B$ introduced by
Theorem~\ref{the:asymptotic}
were respectively equal to $\tilde Q_1^B$ and $\tilde Q_1^A$.

\noindent {\sl iii)} The final value of $d^A, d^B$ is given by 
$d^A(Q_1^A,Q_1^B)$, $d^B(Q_1^A,Q_1^B)$, the value
associated with
$(Q_1^A,Q_1^B)$ the argument of the minimum power transmitted by the network:
$$
(Q_1^A,Q_1^B) = \arg\min_{(\tilde Q_1^A,\tilde Q_1^B)} Q_T(\tilde Q_1^A,\tilde
Q_1^B)\:.
$$

\noindent {\sl iv)} Finally, we choose 
\begin{center}
$d_{\textrm{subopt}}^A=d^A$ and $d_{\textrm{subopt}}^B=d^B$.
\end{center}
Note that the same procedure provides as a byproduct the limit $Q_T$ of the
total transmit power as $Q_T=Q_T(Q_1^A,Q_1^B)$.
\medskip

\noindent {\bf Comments}

It is clear from our previous discussion that the above procedure for
computing $(d^A,d^B)$ can be done in advance prior to resource allocation.
This is essentially due to the fact that the asymptotically optimal pair of
pivot-distances $(d^A,d^B)$ does not depend on the particular cell
configuration, but on an asymptotic or ``average'' state of the network. 
The procedure can be run for instance before base stations are brought into
operation. It can also be done once in a while as the asymptotic distribution
of the users and the average rate requirement $\bar r$ can be subject to
changes: but these changes occur after long periods of time. Therefore, the
number of operations needed for the computation of $(d^A,d^B)$ is not a major
concern because it does not affect the computational complexity of resource
allocation.
%
\subsection{Asymptotic Performance of the Simplified Algorithm}
\label{sec:asymp_subopt}
Denote by $Q_{subopt}\K$ the total power transmitted when our simplified
allocation algorithm is applied. Recall that $Q_T\K$ designates the total power
transmitted by the network when the optimal resource allocation associated with
the joint resource allocation problem (Problem~2 of~\cite{partI}) is used.
\begin{prop}
\label{prop:subopt}
The following equality holds:
\begin{equation*}
\lim_{K\to\infty} Q_{subopt}\K = \lim_{K\to\infty} Q_T\K\;.
\end{equation*}
\end{prop}
Proposition~\ref{prop:subopt} can be proved using the same arguments as the ones
used in Appendix~\ref{app:asym_proof}. The detailed proof is omitted. The above
Proposition
states that the proposed suboptimal algorithm tends to be optimal w.r.t. the
joint resource allocation problem, as the number of users increases.
Therefore, our algorithm is at the same time much simpler than the initial
optimal resource allocation algorithm of~\cite{partI},
and has similar performance at least for a sufficient number of users in each
cell. Section~\ref{sec:simus} will furthermore indicate that even for a moderate
number
of users, our suboptimal algorithm is actually nearly optimal.
\section{On the Convergence of the Simplified Allocation Algorithm}
\label{sec:subopt_feasibility}
As stated before, the simplified algorithm performs the resource
allocation in each Cell~$c$ independently for the protected ${\cal K}_P^c$ and
the non protected ${\cal K}_I^c$ users, which are separated by the predefined
pivot-distance $d_{\textrm{subopt}}^c$.
Resource allocation for the non protected users is done by the
iterative and distributed ping-pong algorithm described in
Section~\ref{sec:sub_optimal_1d}. It was stated in
Section~\ref{sec:sub_optimal_1d} that the convergence of the ping-pong
algorithm is ensured by the feasibility of the the problem of resource
allocation for the non protected users $\{{\cal K}_I^A,{\cal K}_I^B\}$
(Problem~\ref{prob:opt_multi}). If Problem~\ref{prob:opt_multi} is feasible, the
ping-pong algorithm converges. If Problem~\ref{prob:opt_multi} is infeasible,
the ping-pong algorithm diverges. It was also
stated in Section~\ref{sec:sub_optimal_1d} that Problem~\ref{prob:opt_multi}
may not be feasible if arbitrary values of the pivot-distances
$d_{\textrm{subopt}}^A$ and $d_{\textrm{subopt}}^B$ are used. Fortunately,
feasibility of the latter problem will not be an issue if the value of
$d_{\textrm{subopt}}^A$ and $d_{\textrm{subopt}}^B$ are relevantly chosen as
described by the procedure
introduced in Section~\ref{sec:asymp_subopt}. Indeed, it can be shown in this
case that at least for large $K$, the set ${\cal K}_I^c$ will contain the users
who would anyway have been restricted to the interference band $\cal I$ if the
optimal resource allocation of Part~I~\cite{partI} was used. 
More precisely, it can be shown that there exists a value $K_0$ of $K$ beyond
which Problem~\ref{prob:opt_multi} is always feasible. 
The proof of this statement is provided in~\cite{rapport}. It is based on
sensitivity analysis of perturbed optimization problems~\cite{perturbed}.
It is worth mentioning that in our simulations, Problem~\ref{prob:opt_multi}
was feasible in almost all the settings of the system, even for
a moderate number of users per cell as small as 25.
\section{Selection of the Best Reuse Factor}
\label{sec:asym_alpha}
The selection of a relevant value $\alpha$ allowing to optimize the network
performance is of crucial importance as far as cellular network design is
concerned. The definition of an \emph{optimal} reuse factor requires however
some care. The first intuition would consist in searching for the value of
$\alpha$ which minimizes the total power $Q_T\K=Q_T\K(\alpha)$ transmitted by
the network, for a finite number of users~$K$. However, $Q_T\K(\alpha)$ depends
on the particular target rates and the particular positions of users. In
practice, the reuse factor should be fixed prior to the resource allocation
process and its value should be independent of the particular cells
configurations.
A solution adopted by several works in the literature consists in performing
system level simulations and choosing the corresponding value of $\alpha$ that
results in the best average performance. In this context, we cite
\cite{sim_reuse}, \cite{sim_reuse2} and \cite{sim_reuse3} without being
exclusive. In this paper, we are interested in providing analytical methods that
permit to choose a relevant value of the reuse factor. This is why we propose to
select the value $\alpha_{\textrm{opt}}$ of the reuse factor as 
$$
\alpha_{\textrm{opt}} = \arg\min_\alpha \lim_{K\to\infty} Q_T\K(\alpha)\:.
$$
Recall that the limiting power $Q_T = \lim_{K\to\infty} Q_T\K$ is given by
equation~(\ref{eq:limitQT}). In practice, we propose to compute the value of
$Q_T=Q_T(\alpha)$ for several values of $\alpha$ on a grid in the interval
$[0,1]$. For each value of $\alpha$ on the grid, $Q_T(\alpha)$ can be obtained
using the procedure presented in subsection~\ref{sec:QT}. Note also that
complexity issues are of few importance, as the optimization is done prior to
the resource allocation process. It does not affect the complexity of the global
resource allocation procedure.
We shall see in Section~\ref{sec:simus} that significant gains are obtained when
using the optimized value of the reuse factor instead of an arbitrary value.
\section{Simulations}\label{sec:simus}

We first begin by presenting the technical parameters of the system model.
In our simulations, we considered a Free Space Loss model (FSL) characterized by
a path loss exponent $s=2$ as well as the so-called Okumura-Hata (O-H) model for
open areas \cite{okumura} with a path loss exponent $s=3$. 
The carrier frequency is $f_0=2.4 GHz$. At this frequency, path loss in dB 
is given by $\rho_{dB}(x) = 20\log_{10}(x) + 100.04$ in the case where $s=2$,
where $x$ is the distance in kilometers between the BS and the user. In the case
$s=3$, $\rho_{dB}(x) = 30\log_{10}(x) + 97.52$. The signal bandwidth $B$ is
equal to $5$ MHz and the thermal noise power spectral density is equal to 
$N_{0} = -170$ dBm/Hz. Each cell has a radius $D = 500$m.

\noindent
{\bf Asymptotically optimal pivot-distance and frequency reuse factor:}
\begin{figure}[h]
   \begin{minipage}[b]{0.50\linewidth}
      \centering \includegraphics[scale=0.25]{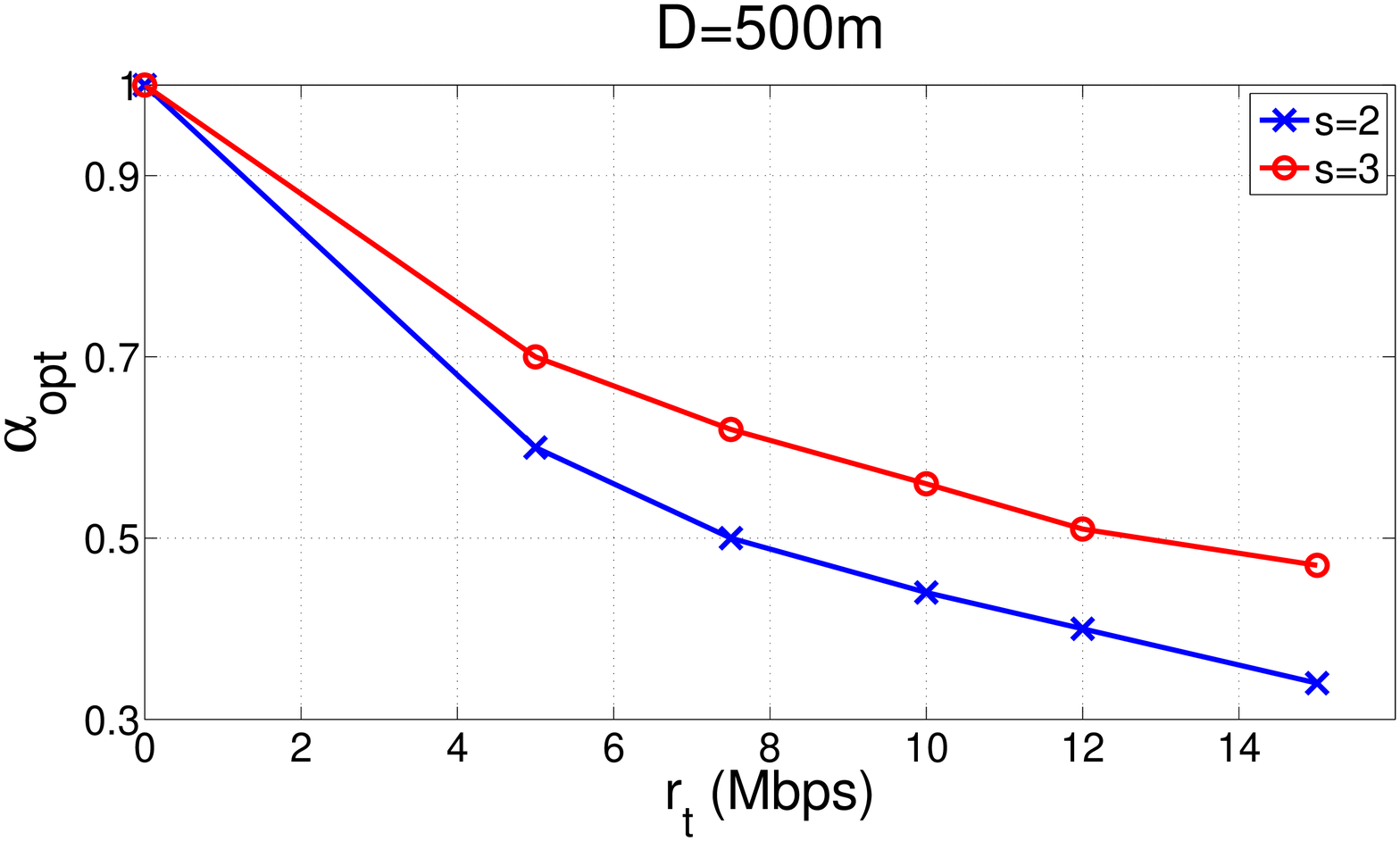}
      \caption{Optimal reuse factor vs. sum rate}
      \label{fig:asym_alpha}
   \end{minipage}\hfill
   \begin{minipage}[b]{0.50\linewidth}   
      \centering \includegraphics[scale=0.25]{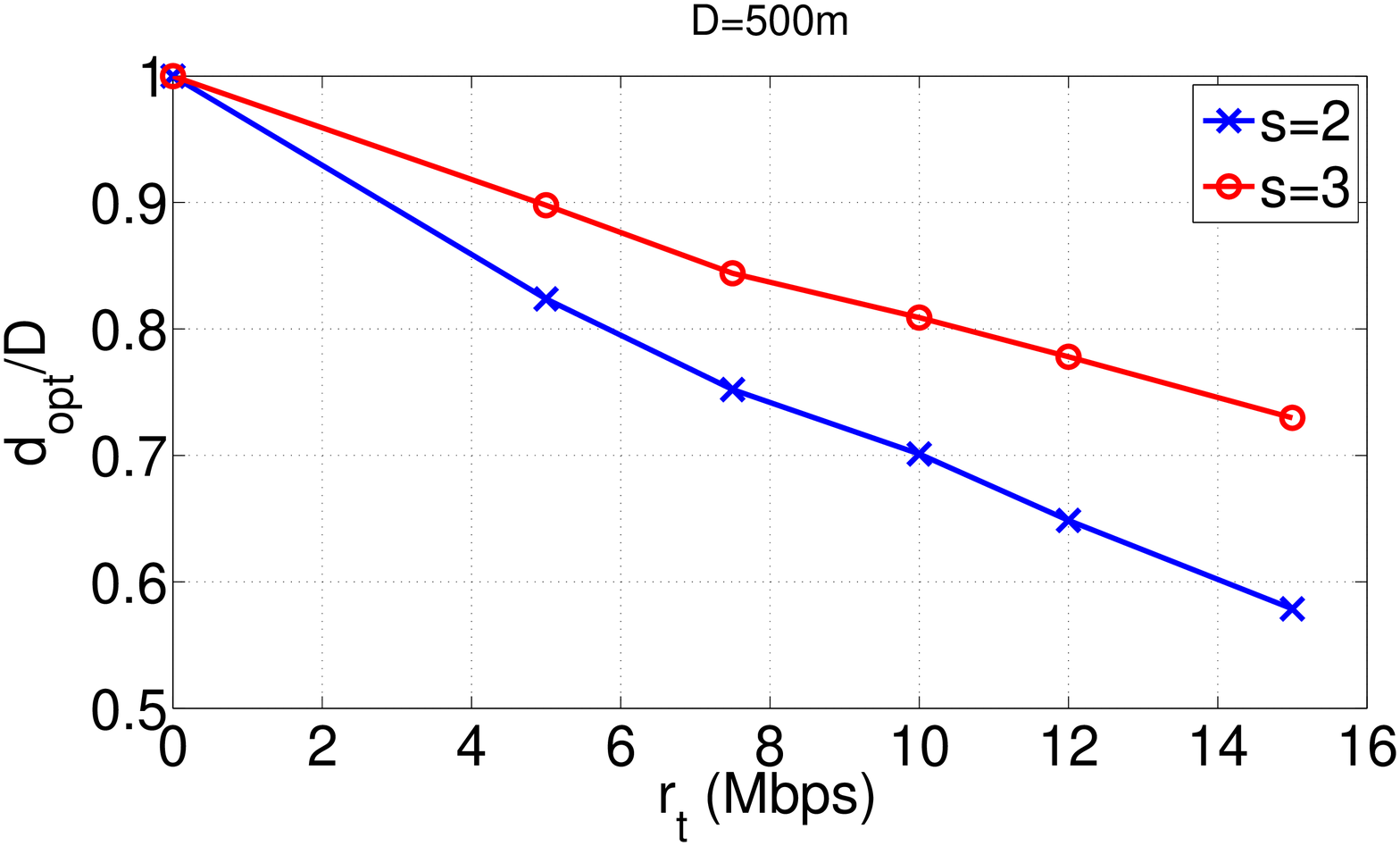}
      \caption{Optimal pivot-distance vs. sum rate}
      \label{fig:asym_d}
   \end{minipage}
\end{figure}
We first apply the results of of Sections~\ref{sec:asymp}
and~\ref{sec:asym_alpha} in order to obtain the values of the asymptotically
optimal pivot-distances $d^A$, $d^B$ and the asymptotically optimal reuse factor
$\alpha_{\textrm{opt}}$. These values are necessary for the implementation of
the simplified allocation algorithm proposed in
Section~\ref{sec:sub_optimal_1d}. Each of the two cells is assumed to have in
the asymptotic
regime the same uniform distribution of users:
$\lambda^A=\lambda^B=\lambda$ where $d\lambda(x)= dx /D$. The average rate
requirement in each cell is assumed to be the same, too:
$\bar{r}^A=\bar{r}^B=\bar{r}$, where $\bar{r}^c$ is defined in
Subsection~\ref{sec:tools} as the average data rate in Cell~$c$ measured in
bits/sec/Hz. In this case, the optimal pivot-distance is the same in each cell
\emph{i.e.,} $d^A=d^B$. Define $d_{\textrm{opt}}=d^A=d^B$. The value of
$d_{\textrm{opt}}$ and
$\alpha_{\textrm{opt}}$ was obtained using the method depicted by
Subsection~\ref{sec:QT}
and Section~\ref{sec:asym_alpha} respectively. Denote
by $r_t$ the total data rate of all the users of a sector measured in bits/sec
($r_t=\bar{r}*B$). Figure~\ref{fig:asym_alpha} and Figure~\ref{fig:asym_d} plot
respectively $\alpha_{\textrm{opt}}$ and the normalized pivot-distance
$d_{\textrm{opt}}/D$ as
functions of the total rate $r_t$ for two values of the path loss exponent:
$s=2$ and $s=3$. Note from Figure~\ref{fig:asym_alpha} that
$\alpha_{\textrm{opt}}$ and
$d_{\textrm{opt}}$ are both decreasing functions of $r_t$. This result is
expected, given
that higher values of $r_t$ will lead to higher transmit powers 
and consequently to higher levels of
interference. More users will need thus to be ``protected'' from the higher
interference. For that purpose, the pivot-position must be closer to the base
station and a larger part of the available bandwidth must be reserved for the
protected bands~$\mathcal{P}_A$ and $\mathcal{P}_B$. Note also that, in the
case $s=3$, ``less protection'' is needed than in the case where $s=2$. In
other words, $d_{\textrm{opt}}(s=3)>d_{\textrm{opt}}(s=2)$ and
$\alpha_{\textrm{opt}}(s=3)>\alpha_{\textrm{opt}}(s=2)$. This observation can be
explained by the
fact that, when the path loss exponent is higher, the interference produced by
the adjacent base station will undergo more fading than in the case when the
path loss exponent is lower.

\begin{figure}[h]
\centering \includegraphics[width=9cm]{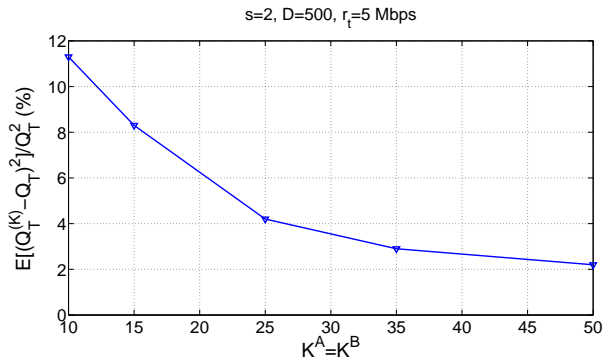}
      \caption{$\frac{\mathbb{E}_{\mathbf{x}}(Q_T\K(\mathbf{x})-Q_T)^2}{Q_T^2}$
               vs. number of users per cell}
      \label{fig:asym_error}
\end{figure}

\noindent
{\bf Simplified resource allocation:}
In Section~\ref{sec:sub_optimal_1d}, we proposed a suboptimal allocation
algorithm characterized by its reduced computational complexity compared to the
optimal allocation algorithm depicted in~\cite{partI}. This algorithm assumes
fixed pivot-distances $d_{\textrm{subopt}}^A$, $d_{\textrm{subopt}}^B$.
Here, we study the performance of this algorithm when $d_{\textrm{subopt}}^A$
and $d_{\textrm{subopt}}^B$ are chosen according to the procedure provided in
Section~\ref{sec:QT} \emph{i.e.,} $d_{\textrm{subopt}}^A=d_{\textrm{opt}}$ and
$d_{\textrm{subopt}}^B=d_{\textrm{opt}}$, where $d_{\textrm{opt}}$ is the
asymptotically optimal pivot-distance defined earlier in this section.
In order to study the performance of this algorithm, we need to compare, for a
large number of system settings, $Q_{\textrm{subopt}}\K$ the total transmit
power that must be spent when applying the simplified algorithm, with $Q_T\K$ 
the total transmit power that must be spent when the optimal resource allocation
scheme of Part~I~\cite{partI} is applied. The results must then be
averaged in order to obtain performance measurements that are independent of the
particular system setting. We consider therefore that users in each cell
are randomly distributed and that the distance separating each user from the
base station is a random variable with a uniform distribution on the interval
$[0,D]$. On the other hand, we assume without restriction that all users have
the same target rate, and that the number of users is the same for the two cells
$K^A=K^B$. 
Define $\mathbf{x}$ as the vector containing the positions 
of all the users in the system \emph{i.e.,} $\mathbf{x}=(x_1$, $x_2$,
$\ldots,x_{K^c})_{c=A,B}$. Recall that $\forall k$, $x_k$ is a random variable
with a uniform distribution on $[0,D]$. For each realization of $\mathbf{x}$,
define $Q_T\K(\mathbf{x},\alpha)$ as the total transmit power
that results from applying the optimal joint resource allocation scheme of
Part~I with the value of the reuse factor fixed to
$\alpha$. Define $Q_T\K(\mathbf{x})=\min_{\alpha}Q_T\K(\mathbf{x},\alpha)$. In
the same way, denote by $Q_{\textrm{subopt}}\K(\mathbf{x})$ the total transmit
power that
results from applying the simplified resource allocation scheme of
Section~\ref{sec:sub_optimal_1d} with the value of the reuse factor fixed to
$\alpha_{\textrm{opt}}$ defined in Section~\ref{sec:asym_alpha}. For each
realization of the random vector $\mathbf{x}$, the values of $Q_T\K(\mathbf{x})$
and $Q_{\textrm{subopt}}\K(\mathbf{x})$ were calculated and then averaged to
obtain $\mathbb{E}_{\mathbf{x}}[Q_T\K(\mathbf{x})]$ and
$\mathbb{E}_{\mathbf{x}}[Q_{\textrm{subopt}}\K(\mathbf{x})]$ respectively.
\begin{figure}[h]
   \begin{minipage}[b]{0.49\linewidth}
      \centering \includegraphics[scale=0.25]{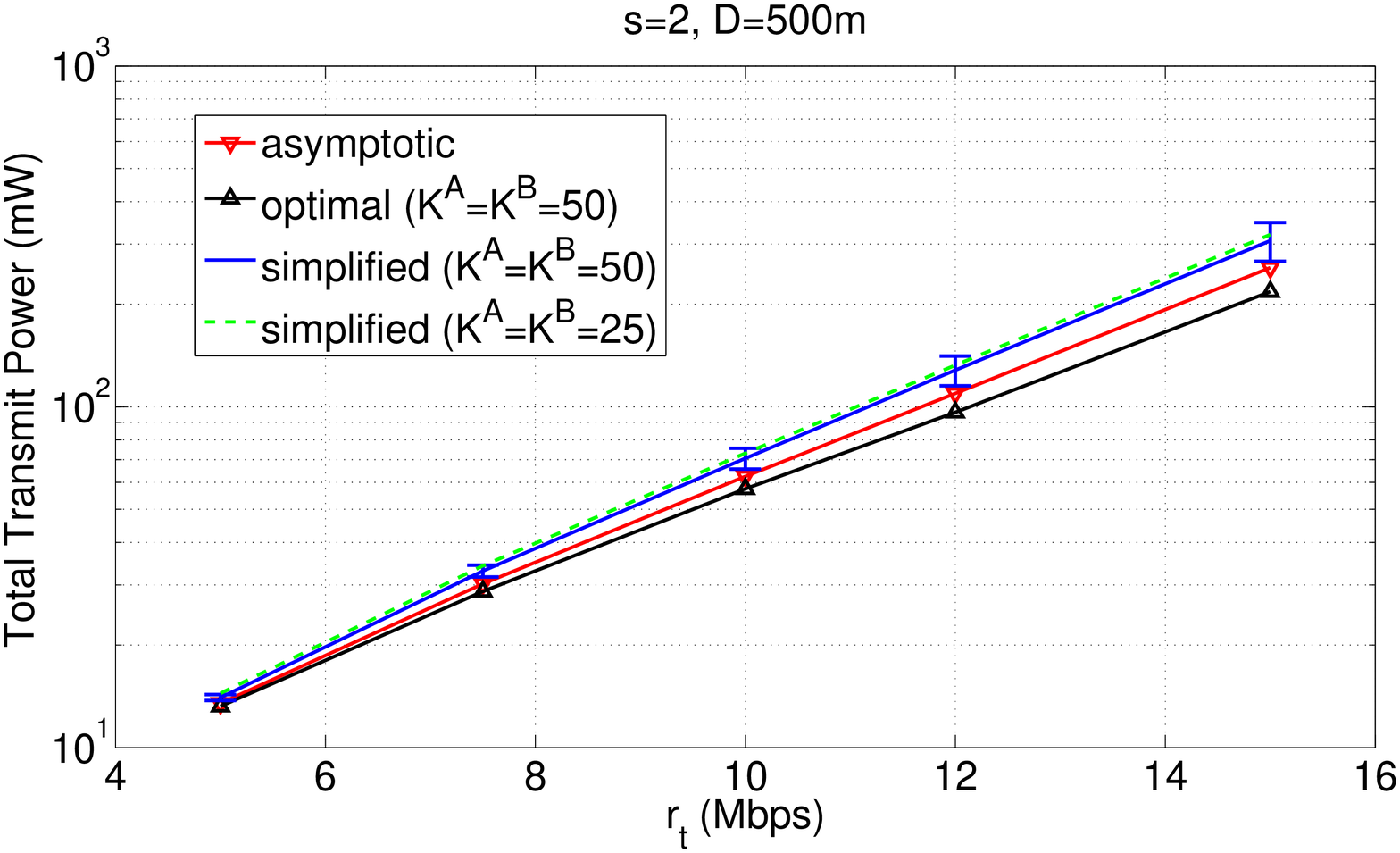}
      \caption{Optimal and suboptimal transmit power vs. sum rate}
      \label{fig:sous-optimal-50}
   \end{minipage}\hfill
   \begin{minipage}[b]{0.49\linewidth}   
      \centering
\includegraphics[scale=0.25]{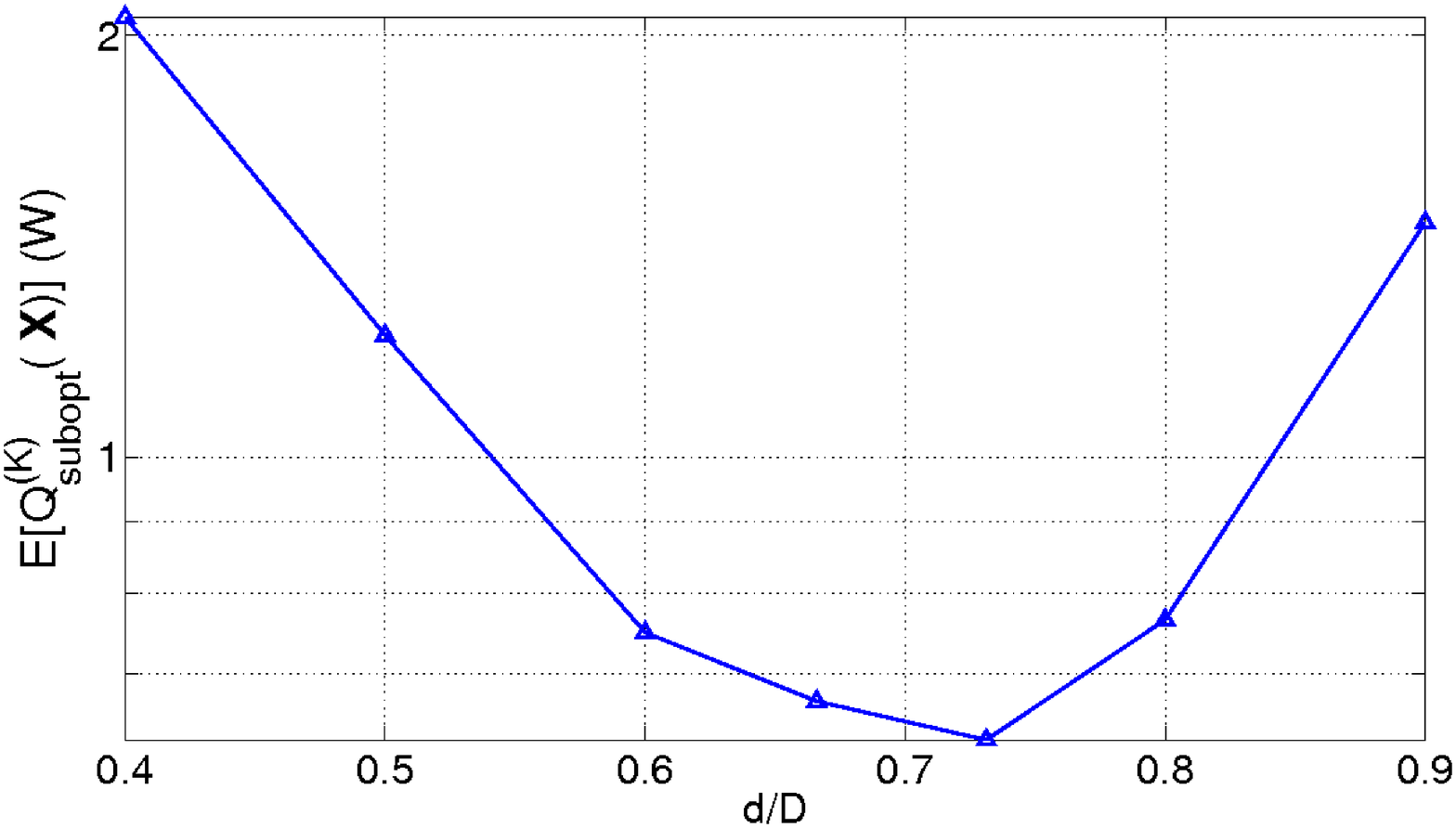}
      \caption{Transmit power vs. the pivot-distance $d$ for the simplified
               allocation scheme ($r_t=10 Mbps$, $K^c=50$)}
      \label{fig:subopt_sensitivity}
   \end{minipage}
\end{figure}
We plot in Figure~\ref{fig:sous-optimal-50} the values of
$\mathbb{E}_{\mathbf{x}}[Q_T\K(\mathbf{x})]$ and
$\mathbb{E}_{\mathbf{x}}[Q_{\textrm{subopt}}\K(\mathbf{x})]$ for a range of
values of the sum rate $r_t$ measured in bits/sec 
($r_t=\sum_{k=1}^{K^c}R_k B$) in two cases: $K^c=25$ and $K^c=50$. 
The error bars in the figure represents the variance of the random variable
$Q_{\textrm{subopt}}\K(\mathbf{x})$
in the case $K^c=50$. In the same figure, the corresponding values of the
asymptotic transmit power $Q_T$ defined by Theorem~\ref{the:asymptotic} are also
plotted. This figure shows that, even for a reasonable number of users equal to
$25$ in each cell, the transmit power needed when we apply the suboptimal
algorithm is very close to the power needed when we apply the optimal resource
allocation scheme. 
The gap between the two powers is of course even smaller for
$K^c=50$. This result validates Proposition~\ref{prop:subopt} which states
that our proposed suboptimal resource allocation scheme is asymptotically
optimal. Figure~\ref{fig:subopt_sensitivity} is dedicated to illustrate the
sensitivity of the simplified allocation scheme with respect to the
pivot-distance $d_{\textrm{subopt}}$ in the case $K^c=50$. For that sake, the
figure plots the total
transmit power resulting from applying the simplified scheme as a function of
$d_{\textrm{subopt}}$. The minimum in the figure corresponds to the
asymptotically optimal pivot
distance $d_{\textrm{subopt}}=d_{\textrm{opt}}$. We note that using values
different from $d_{\textrm{opt}}$ increases
the suboptimality of the simplified scheme.
Let us go back to Figure~\ref{fig:sous-optimal-50}. The latter figure shows 
that over the range of the considered values of the total data rate $r_t$, the
total transmit power $\mathbb{E}_{\mathbf{x}}[Q_T\K(\mathbf{x})]$ for
$K^c=50$ is practically equal to the asymptotic power $Q_T$. This result
suggests that, for a number of users equal to $50$ in each cell, the system is
already in its asymptotic regime. In order to validate the latter affirmation,
one still needs to investigate the value of the mean square error
$(Q_T\K-Q_T)^2$ as well. This is done by Figure~\ref{fig:asym_error} which plots
$\frac{\mathbb{E}_{\mathbf{x}}(Q_T\K(\mathbf{x})-Q_T)^2}{Q_T^2}$, the mean
square error normalized by $Q_T^2$.
\section{Conclusions}
In this pair of papers, the resource allocation problem for sectorized downlink
OFDMA systems has been studied in the context of a partial reuse factor
$\alpha\in [0,1]$. In the first part of this work, the general solution to the
(nonconvex) optimization problem has been provided. It has been proved that the
solution admits a simple form and that the initial tedious problem reduces to
the identification of a limited number of parameters. As a noticeable property, 
it has been proved that the optimal resource allocation policy is ``binary'':
there exists a pivot-distance to the BS such that users who are farther than
this distance should only modulate protected subcarriers, while closest users
should only modulate reused subcarriers. A resource allocation algorithm has
been also proposed.

In the second part, we proposed a suboptimal resource allocation algorithm
which avoids the costly search for parameters such as the optimal
pivot-distance. In the proposed procedure, the optimal pivot-distance is simply
replaced by a fixed value. In order to provide a method to relevantly select
this fixed pivot-distance, the asymptotic behavior of the optimal resource
allocation has been studied as the number of users tends to infinity.
In the case where the fixed pivot-distance associated with the
simplified algorithm is chosen to be equal to the asymptotically
optimal pivot-distance, it has been shown that our simplified resource
allocation
algorithm is asymptotically equivalent to the optimal one as the number of users
increases. Simulations proved the relevancy of our algorithm even for a small
number of users. Using the results of the asymptotic study,  
the optimal value of the reuse factor has been characterized. 
It is defined as the value of $\alpha$ which minimizes the asymptotic value of
the minimum transmit power. Our simulations proved that substantial
improvements in terms of spectral efficiency can be expected when using the
relevant value of the reuse factor.
\appendices
\section{Proof of Theorem~\ref{the:asymptotic}}
\label{app:asym_proof}

Theorem~\ref{the:asymptotic} characterizes the asymptotic behaviour of the
minimal transmit power resulting from applying the optimal resource allocation
when the number of users $K$ tends to infinity.
It is thus useful at this point to recall the theorem given in the first part 
of this work which characterizes the optimal allocation for finite values of
$K$. Define the function $F(x) = \EE{\frac{Z}{1+f^{-1}(x)Z}}$. For each cell
$c=A, B$
and for each $l=1\dots K^c$, define by $a_{l}^c$ and $b_{l}^c$ 
the unique positive numbers such that
$\sum_{k=1}^{l}\frac{R_{k}}{C\left(g_{k,1} a_l^c\right)}=\alpha$ and
$\sum_{k=l+1}^{K^c}\frac{R_{k}}{C(g_{k,2}b_l^c)}=\frac{1-\alpha}{2}$,
with $a_0^c=b_{K^c}^c=0$ by convention.
\begin{theo}[\cite{partI}]
\label{the:multi}
~\\{\bf(A)} Any global solution to the joint
resource allocation problem has the following form. For each
Cell~$c$, there exists an integer $L^c\in\{1,\dots, K^c\}$, and there exist four
positive numbers $\beta_1^c$, $\beta_2^c$, $\xi^c$, $Q_1^{\bar c}$  such that 
\begin{enumerate}
\item For each $k<L^c$,
\begin{equation}
\label{eq:multiinf}
	\begin{array}[h]{l|l}
	 \dsp P\ku^c={g\ku(Q_1^{\bar c})}^{-1}f^{-1}\left(\frac{g\ku(Q_1^{\bar
c})}{1+\xi^c}\beta_1^c\right) & P\kd^c=0 \\
	\dsp \gamma\ku^c=\frac{R_k}{\EE{\log\left(1+{g\ku(Q_1^{\bar c})} P\ku^c
Z\right)}} & \gamma\kd^c=0
	\end{array}
\end{equation}

\item For each $k>L^c$,
\begin{equation}
\label{eq:multisup}
	\begin{array}[h]{l|l}
	 P\ku^c=0 & \dsp P\kd^c=g\kd^{-1}f^{-1}(g\kd \beta_2^c) \\
	\gamma\ku^c=0 & \dsp \gamma\kd^c=\frac{R_k}{\EE{\log\left(1+g\kd P\kd^c
Z\right)}}
	\end{array}
\end{equation}

\item For $k=L^c$
\begin{equation}
\label{eq:multiL}
	\begin{array}[h]{l|l}
	 \dsp P\ku^c={g\ku(Q_1^{\bar c})}^{-1}f^{-1}\left(\frac{g\ku(Q_1^{\bar
c})}{1+\xi^c} \beta_1\right) & 
	 \dsp P\kd^c=g\kd^{-1}f^{-1}(g\kd \beta_2^c) \\
	 \dsp \gamma\ku^c=\alpha-\sum_{l=1}^{k-1}\gamma\lu^c &
	 \dsp \gamma\kd^c=\frac{1-\alpha}{2}-\sum_{l=k+1}^{K^c}\gamma\ld^c.
	\end{array}
\end{equation} 
\end{enumerate}
{\bf(B)} For each $c=A,B$, the system ${\cal S}^c(Q_1^A,Q_1^B)$ formed by the
following four equations is satisfied.
\begin{gather}
  L^c = \min \left\{l=1\dots K^c \big/ \frac{g\lu(Q_1^{\bar c})}{1+\xi^c}
  F\left(\frac{g\lu(Q_1^{\bar c})}{1+\xi^c}a_{l}\right) \leq
g_{l,2}F\left(g_{l,2}b_l\right)\right\} 
\label{eq:Lmulti}\\
  \frac{g_{L^c,1}(Q_1^{\bar c})}{1+\xi^c} 
  F\left(\frac{g_{L^c,1}(Q_1^{\bar c})}{1+\xi^c} \beta_1^c\right)=g_{L^c,2}
F(g_{L^c,2}\beta_2^c) 
\label{eq:FFmulti}\\
  \gamma_{L^c,1}^c C\left(\frac{g_{L^c,1}(Q_1^{\bar
c})}{1+\xi^c}\beta_1^c\right)+
  \gamma_{L^c,2}^c C(g_{L,2}{\beta}_{2}^c)=R_{L^c}
\label{eq:RLCLmulti}\\
	\sum_k^{L^c}\gamma\ku^c P\ku^c= Q_1^c\:,
\label{eq:Qmulti}
\end{gather}
where the values of $\gamma\ku^c$ and $P\ku^c$ in~(\ref{eq:Qmulti}) are the
functions of $(\beta_1^c,\beta_2^c,\xi^c)$ defined by
equation~(\ref{eq:multiinf}).\\
{\bf(C)} Furthermore, for each $c=A,B$ and for any arbitrary values
$\tilde{Q}_1^A$ and $\tilde{Q}_1^B$, 
the system of equations ${\cal S}^c(\tilde{Q}_1^A,\tilde{Q}_1^B)$ admits at most
one solution $(L^c,\beta_1^c,\beta_2^c,\xi^c)$.
\end{theo}
In subsection~\ref{sec:tools}, we obtained that for each cell $c=A,B$, 
\begin{eqnarray}
  Q_1\cK &=& \frac{K^c}{B} \int\!\!\!\! \int_{\Delta_1\cK} r{\cal F}(x,\beta_1\cK,Q_1^{\bar{c},(K)},\xi\cK)d\nu\cK(r,x)+ 
  o_K(1)\label{eq:fonctionMesureQ1}\\
  Q_2\cK &=& \frac{K^c}{B} \int\!\!\!\! 
  \int_{\Delta_2\cK} r{\cal F}(x,\beta_2\cK,0,0)d\nu\cK(r,x)+ o_K(1)
  \label{eq:fonctionMesureQ2}
\end{eqnarray}
where $\Delta_1\cK=[0,\rho]\times [\epsilon, d\cK]$ and
$\Delta_2\cK=[0,\rho]\times [d\cK, D]$ and where
$d\cK$ is the pivot-distance \emph{i.e.}, the position of user $L\cK$.
Our aim is to prove that $Q_T\K = \sum_c Q_1\cK + Q_2\cK$ converges as
$K\to\infty$, and
to characterize the limit.
For each cell $c\in\{A,B\}$, sequence $d\cK$ is bounded by definition 
($d\cK \in [0,D]$). Consider a subsequence $\phi_K$ such that
$(d^{A,(\phi_K)},d^{B,(\phi_K)})$ converges to a certain
limit, say $(d^A,d^B)$. We prove that in this case, all quantities 
$Q_1\cpK$, $Q_2\cpK$, $\beta_1\cpK$, $\beta_2\cpK$, $\xi\cpK$ converge to some
values $Q_1^c$, $Q_2^c$, $\beta_1^c$, $\beta_2^c$, $\xi^c$
which we shall characterize. Focus for instance on sequence $\beta_2\cpK$. 
Recalling that $\gamma_{L\cK,2}^c$ tends to zero as $K\to\infty$
($\gamma_{L\cK,2}^c=o_K(1)$)
and replacing each $\gamma\kd^c$ with
expression~(\ref{eq:multiinf})  $\gamma\kd^c=R_k/C(g\kd^c \beta_2\cK)$, we
obtain immediately
\begin{equation}
  \frac{1}{B}\sum_{k>L\cK} r_k {\cal
G}(x,\beta_2\cK,0,0)+o_K(1)=\frac{1-\alpha}{2}\:,
\label{eq:intermC3bis}
\end{equation}
where we defined 
\begin{equation}\label{eq:calG}
{\cal G}(x,\beta,{\cal Q},\xi)=\frac{1}{C\left(\frac{g_1(x,{\cal
Q})}{1+\xi}\beta\right)}
\end{equation}
for each $x,\beta, {\cal Q}, \xi$.
In the asymptotic regime, we obtain the following lemma.
\begin{lemma}
  \label{lem:betad}
As $K\to\infty$, sequence $\beta_2\cpK$ converges to the unique solution
$\beta_2^c$ to the following equation:
\begin{equation}
  \frac{t}{2}\int_0^{\rho}\int_{{d}^c}^D r {\cal G}(x,\beta_2^c,0,0)d\nu^c(r,x)=\frac{1-\alpha}{2}\;.
\label{eq:eqbetad}
\end{equation}
\end{lemma}
\begin{proof}
 Existence and uniqueness of the solution to~(\ref{eq:eqbetad}) is straightforward since
 function~$\beta\mapsto{\cal G}(x,\beta,{\cal Q},\xi)$ is strictly decreasing from $\infty$ to 0 
 on $\mathbb{R}_+$. We remark that sequence $\beta_2\cpK$ is bounded \emph{i.e.},
 $\beta_2\cpK\leq \kappa$ for a certain constant $\kappa$. In order to prove this claim,
 assume that there exists a subsequence $\beta_2^{c,\phi_{\zeta(K)}}$ which converges to
 infinity. This hypothesis implies that the subsequence given by the lhs
 of~(\ref{eq:intermC3bis}) for $K$ of the form $K=\zeta(K')$ converges to zero as $K'\to\infty$.
 This is in contradiction with~(\ref{eq:intermC3bis}) which states that the 
 latter sequence converges to $\frac{1-\alpha}{2}$. 
 Using similar arguments, it can be shown that $\beta_2\cpK$ is lower bounded by a certain $\epsilon'>0$
 \emph{i.e.}, $\epsilon' < \beta_2\cpK <\kappa$.
 Denote by $\beta_2$ any accumulation point of $\beta_2\cpK$ and define $\beta_2\ctK$ a subsequence
 of $\beta_2\cpK$ (\emph{i.e.}, $\theta_K$ coincides with $\phi_{\zeta(K)}$ for a certain function $\zeta$)
 which converges to $\beta_2$. We prove that $\beta_2$ is given by~(\ref{eq:eqbetad}). 
 Define $G(r,x,y)=r{\cal G}(x,y,0,0)$. We show that the difference
 \begin{eqnarray*}
 \Delta_K &=& \left| \int_{0}^{\rho}\int_{d\ctK}^D G(r,x,\beta_2\ctK)d\nu\ctK(r,x)- \int_{0}^{\rho}\int_{{d}^c}^D G(r,x,\beta_2\ctK)d\nu^c(r,x)\right|
 \end{eqnarray*}
 tends to zero as $K\to\infty$. By the triangular inequality,
 \begin{eqnarray*}
 \Delta_K &\leq& \left| \int_{0}^{\rho}\int_{d\ctK}^D G(r,x,\beta_2\ctK)d\nu\ctK(r,x)- \int_{0}^{\rho}\int_{{d}^c}^D G(r,x,\beta_2\ctK)d\nu\ctK(r,x)\right|\\
 &&+ \left|\int_{0}^{\rho}\int_{{d}^c}^D G(r,x,\beta_2) d\nu\ctK(r,x)-  \int_{0}^{\rho}\int_{{d}^c}^D G(r,x,\beta_2\ctK)d\nu^c(r,x)\right|\\
 &&+ \int_{0}^{\rho}\int_{{d}^c}^D \left|G(r,x,\beta_2\ctK)-G(r,x,\beta_2) \right|d\nu\ctK(u,x)\:.
 \end{eqnarray*}
 Respectively denote by $\Delta_{K,1}$, $\Delta_{K,2}$, $\Delta_{K,3}$ the first, second and third terms of the above equation.
 We first study $\Delta_{K,1}$. Clearly, function $G(u,x,\beta)$ is bounded on
$[0,\rho]\times [\epsilon, D]\times [\epsilon',\kappa]$. 
 Denote by $\xi$ an upper bound. Then, $\Delta_{K,1}\leq \xi \nu_c^{\theta_K}(I_K)$, where 
 $I_K=[0,\rho] \times [d\ctK,{d}^c]$ (or $I_K=[0,\rho] \times [{d}^c,d\ctK]$ 
 if ${d}^c<d\ctK$).
 Recall that $d\ctK$ converges to ${d}^c$ by definition, so that $\nu^c(I_K)=\zeta^c([0,\rho])\lambda^c([d\ctK,{d}^c])$ 
 converges to zero as long as measure $\lambda^c$ has no mass point at ${d}^c$.
 Since $\nu\ctK$ converges weakly to $\nu^c$, 
 it is straightforward to show that $\nu\ctK(I_K)$, 
 and thus $\Delta_{K,1}$, tend to zero.  
 Now focus on $\Delta_{K,2}$. The first term $\int\int G(r,x,\beta_2) d\nu\ctK(r,x)$ 
 which composes $\Delta_{K,2}$
 converges to $\int\int G(r,x,\beta_2) 	d\nu^c(r,x)$ by the weak convergence of $\nu\ctK$ to $\nu^c$.
 The second term $\int\int G(r,x,\beta_2\ctK) d\nu^c(r,x)$ 
 converges to the same limit by Lebesgue's dominated convergence
 Theorem. Thus, $\Delta_{K,2}$ tends to zero. In order to prove that $\Delta_{K,3}$ tends to zero, we remark that 
 $\sup \left|\dron{G(x,r,\beta)}{\beta}\right|<\infty$, where the supremum is taken w.r.t. $(x,r,\beta)\in 
 [0,\rho]\times [\epsilon, D]\times [\epsilon',\kappa]$. Denote by $C$ the
latter supremum. We easily obtain
 $\left|G(r,x,\beta_2\ctK)-G(r,x,\beta_2) \right|\leq C  \left|\beta_2\ctK-\beta_2 \right|$, so that
 $\Delta_{K,3}\leq C  \left|\beta_2\ctK-\beta_2 \right|
\nu_c^{\theta_K}([0,\rho]\times [{d}^c, D])$.
 Since $\nu_c^{\theta_K}$ is a probability measure, $\Delta_{K,3}\leq C  \left|\beta_2\ctK-\beta_2 \right|$.
 Thus $\Delta_{K,3}$ tends to zero as $K$ tends to infinity. Putting all pieces together, $\Delta_K$ tends to zero.
 Using~(\ref{eq:intermC3bis}), 
 $\frac{t}{2} \int_{0}^{\rho}\int_{{d}^c}^D G(r,x,\beta_2\ctK)d\nu^c(r,x)$ converges to
 $\frac{1-\alpha}{2}$. By continuity arguments, $\beta_2=\lim_K \beta_2\ctK$ satisfies~(\ref{eq:eqbetad}).
 Thus $\beta_2\cpK$ is a bounded sequence such that any accumulation point is equal to $\beta_2$
 defined by~(\ref{eq:eqbetad}). Thus $\lim_K \beta_2\cpK=\beta_2$.
 \end{proof}
 Using Lemma~\ref{lem:betad}, we may now characterize the limit of~(\ref{eq:fonctionMesureQ2})
 as $K\to\infty$. Using the fact that\\ 
$\lim_K \beta_2\cpK=\beta_2^c$ and $\lim_K d\cpK={d}^c$ 
 along with some technical arguments which are similar to the ones used in the 
 proof of Lemma~\ref{lem:betad}, we obtain 
 \begin{eqnarray}
 Q_2\cpK&=&\frac{K^c}{B}\int_{0}^{\rho}\int_{{d}^c}^D r{\cal F}(x,\beta_2^c,0,0)d\nu\cpK(r,x)+o_K(1)
 \end{eqnarray}
 where $\beta_2^c$ is the unique solution to~(\ref{eq:eqbetad}). 
  As $\nu\cpK$ converges weakly to $\nu^c$, $Q_2\cpK$ converges to 
  \begin{equation}\label{eq:Q2_beta2_d}
  Q_2^c=  \frac{t}{2}\int_{0}^{\rho}\int_{{d}^c}^Dr{\cal F}(x,\beta_2,0,0)d\nu^c(r,x) \:.
  \end{equation}
The same approach can be used to analyze the behavior of sequences
$Q_1\cpK$ and $\beta_1\cpK$ for each $c=A,B$. After similar derivations, we obtain
the following result.
As $K\to\infty$, sequence ($\beta_1\ApK$, $Q_1\ApK$, $\xi\ApK$, $\beta_1\BpK$, $Q_1\BpK$, $\xi\BpK$) 
converges to the \emph{unique} solution $(\beta_1^A,Q_1^A,\xi^A,\beta_1^B,Q_1^B,\xi^B)$ 
to the following system of six equations:
\begin{equation}
  \label{eq:gamma1_beta1_d}
  \left.\begin{array}{l}
    \dsp Q_1^c= \frac{t}{2}\int_{0}^{\rho}\int_\epsilon^{{d}^c}r{\cal F}(x,\beta_1^c,Q_1^{\bar c},\xi^c)\:d\nu^c(r,x)\\
    \dsp \frac{t}{2}\int_{0}^{\rho}\int_{\epsilon}^{{d}^c}r{\cal G}(x,\beta_1^c,Q_1^{\bar c},\xi^c)\nu^c(r,x)=\alpha\\
    \dsp \frac{g_1(d^c,Q_1^{\bar c})}{1+\xi^c} F\left(\frac{g_1(d^c,Q_1^{\bar c})}{1+\xi^c} \beta_1^c\right)=
    g_2(d^c) F(g_2(d^c)\beta_2^c)
  \end{array}\;\;\right\}\;\;\;\; c=A,B\:,
\end{equation}
where $\beta_2^c$ and $d^c$ are the limits of $\beta_2\cpK$ and $d\cpK$ respectively.
We discuss now the existence and the uniqueness of the solution to the above system of equation. 
For that sake, recall the definition of functions ${\cal F}$ and $\cal G$ given by
\eqref{eq:calF} and \eqref{eq:calG} respectively. Note that 
${\cal F}(x,\beta,{\cal Q},\xi)={\cal F}(x,\frac{\beta}{1+\xi},{\cal Q},0)$, and
that ${\cal G}(x,\beta,{\cal Q},\xi)={\cal G}(x,\frac{\beta}{1+\xi},{\cal Q},0)$.
Define $\tilde{\beta_1^c}=\frac{\beta_1^c}{1+\xi^c}$ for $c\in\{A,B\}$. By applying
this new notation, The first two equations of system~\eqref{eq:gamma1_beta1_d} give place to
the following system of four equations:
\begin{equation}
  \label{eq:gamma1_beta1}
  \left.\begin{array}{l}
    \dsp Q_1^c= \frac{t}{2}\int_{0}^{\rho}\int_\epsilon^{{d}^c}r{\cal F}(x,\tilde{\beta_1^c},Q_1^{\bar c},0)\:d\nu^c(r,x)\\
    \dsp \frac{t}{2}\int_{0}^{\rho}\int_{\epsilon}^{{d}^c}r{\cal G}(x,\tilde{\beta_1^c},Q_1^{\bar c},0)\nu^c(r,x)=\alpha
  \end{array}\;\;\right\}\;\;\;\; c=A,B\:.
\end{equation}
The existence and the uniqueness of the solution $(\tilde{\beta_1^c},Q_1^c)_{c=A,B}$ 
to the system~\eqref{eq:gamma1_beta1} was thoroughly studied
in~\cite{gau-hac-cib-1}. Applying the results of~\cite{gau-hac-cib-1} in our
context, we conclude that 
$(\frac{\beta_1^c}{1+\xi^c},Q_1^c)_{c=A,B}=(\tilde{\beta_1^c},Q_1^c)_{c=A,B}$ 
is unique. 
We turn now back to the third equation of system~\eqref{eq:gamma1_beta1_d} to
get the following
equality $\xi^c=\frac{g_1(d^c,Q_1^{\bar c})
F\left(g_1(d^c,Q_1^{\bar c})\frac{\beta_1^c}{1+\xi^c}\right)}{g_2(d^c) F(g_2(d^c)\beta_2^c)}-1$.
The latter equation proves the uniqueness of $\xi^c$ for $c=A,B$. The uniqueness of $\beta_1^c$ 
follows directly from the same equation.\\
So far, we have proved the uniqueness of the solution to the system~\eqref{eq:gamma1_beta1_d} of equation.
As for the convergence of sequences ($\beta_1\ApK$, $Q_1\ApK$, $\xi\ApK$, $\beta_1\BpK$, 
$Q_1\BpK$, $\xi\BpK$) to this unique solution, its proof is omitted here due to the lack of space, 
but follows the same ideas as the proof of convergence of ($\beta_2\ApK$, $Q_2\ApK$) and 
($\beta_2\BpK$, $Q_2\BpK$) provided above.

So far, me managed to prove that for any convergent subsequence
$(d\ApK,d\BpK)\to (d^A,d^B)$, the set of parameters 
($Q_1\cpK$, $Q_2\cpK$, $\beta_1\cpK$, $\beta_2\cpK$, $\xi\cpK$)$_{c=A,B}$
converges to some values $Q_1^c$, $Q_2^c$, $\beta_1^c$, $\beta_2^c$, $\xi^c$ which are completely characterized by
the system of equations~(\ref{eq:eqbetad}),~(\ref{eq:Q2_beta2_d}) and~(\ref{eq:gamma1_beta1_d}),
as functions $(d^A,d^B)$. 
Using decomposition $\nu^c=\zeta^c \times \lambda^c$, the system formed by 
equations~(\ref{eq:eqbetad}),~(\ref{eq:Q2_beta2_d}) and~(\ref{eq:gamma1_beta1_d}) 
is equivalent to the system \Sinf provided in Theorem~\ref{the:asymptotic}.
At this point, we thus proved that at least for some subsequences $\phi_K$ defined as above,
the subsequence $Q_T\cpK$ converges to a limit which has the form given by Theorem~\ref{the:asymptotic}.
The remaining task is to prove that $Q_T\K$ is a convergent sequence.

First, note that $Q_T\K$ is a bounded sequence. Indeed, $Q_T\K$ is defined
as the minimum power that can be transmitted by the network to satisfy 
the rate requirements. By definition, $Q_T\K$ is thus less than the power
obtained when using the naive solution which consists in forcing each
base station to transmit only in the protected band ($\gamma\ku^c$ is forced to zero
for each user $k$ of each cell $c$). Now it can easily be shown that when $K\to\infty$,
the power associated with this naive solution converges to a constant.
As a consequence, one can determine an upper-bound on $Q_T\K$ which does not
depend on $K$. 

Second, assume for instance that $Q_T$ and $Q_T'$ are two accumulation points
of sequence $Q_T\K$. By contradiction, assume that $Q_T < Q_T'$.
Extract for instance a certain subsequence of $Q_T\K$ which converges to $Q_T$.
Inside this subsequence, one can further extract a subsequence, say $\theta_K$,
such that
\begin{equation*}
 Q_T^{(\theta_K)} \longrightarrow Q_T, \qquad d\ctK \longrightarrow d^c, \;\;
\forall c=A,B
\end{equation*}
where $d^A$ and $d^B$ are some constants both 
(just use the fact that $d\cK$ is bounded for each $c$).
Clearly, $Q_T$ can be written as in~(\ref{eq:limitQT}), where
parameters $\beta_1^c$, $\beta_2^c$, $d^c$, $Q_1^c$, $\xi^c$ satisfy the system of equations \Sinf.
We now consider the following \emph{suboptimal} resource allocation policy
for finite numbers of users $K^A$ and $K^B$. 
In each cell $c\in\{A,B\}$, users $k$ whose distance $x_k$ 
to their BS is less than $d^{c}$ are forced to modulate in
the interference band $\cal I$ only, while users $k$ which are farther than $d^{c}$ are forced
to modulate in the protected band ${\cal P}_c$ only. In other words, for each 
user $k$ in cell $c$, we impose
\begin{equation}
  \label{eq:addconstr}
{\bf [C']}  \left\{\begin{array}{l}
    x_k < d^{c} \;\; \Rightarrow \gamma\kd^c=P\kd^c=0 \\
    x_k \geq d^{c} \;\; \Rightarrow \gamma\ku^c=P\ku^c=0 
  \end{array}\right.\;\; \forall c=A,B \:.
\end{equation}
Particular values of the (nonzero) resource allocation parameters $\gamma\ki^c, P\ki^c$ 
are obtained by minimizing the classical joint multicell resource allocation problem
(Problem~2 in~\cite{partI}),
only including the additional constraint ${\bf [C']}$. As a new constraint has been added,
it is clear that the total power transmitted by the network, say $Q_T^{(K),*}$, is always larger than
the total power $Q_T^{(K)}$ achieved by the optimal resource allocation, for any $K$.
On the other hand, using the same asymptotic tools as previously, 
it can be shown after some algebra that 
$$
\lim_K Q_T^{(\theta_K),*} = \lim_K Q_T^{(\theta_K)} = Q_T\:.
$$
In other words, this suboptimal solution performs as good as the optimal one when $K$
has the form $K=\theta_{K'}$ for some $K'$. Although we omit the proof, this observation
is rather intuitive. Indeed for such $K=\theta_{K'}$, the optimal values of the
pivot-distances converge
to the arbitrary ones $d^A$, $d^B$. Even more importantly, it can be shown that the
total power $Q_T^{(K),*}$ spent when using the suboptimal procedure converges as $K\to\infty$.
Therefore,
$$
\lim_K Q_T^{(K),*} = Q_T\:.
$$
Now consider a subsequence $\psi_K$ such that $\lim_K Q_T^{(\psi_K)} = Q_T' > Q_T$,
and compare our suboptimal allocation policy to the optimal one for the $K$'s of
the form $K=\psi_{K'}$. As $\lim_K Q_T^{(\psi_K)} > \lim_K Q_T^{(\psi_K),*}$, there exist a certain $\epsilon>0$
and there exists a certain $K_0$ such that for any $K>K_0$,
$$
Q_T^{(\psi_K)} > Q_T^{(\psi_K),*}+\epsilon\:.
$$
The above inequality contradicts the fact that $Q_T^{(\psi_K)}$ is the global solution
to the joint multicell resource allocation problem (Problem~2 in~\cite{partI}).
Therefore, $Q_T'$ necessarily coincides with $Q_T$. This proves that $Q_T\K$ converges
to $Q_T$.
To complete the proof of Theorem~\ref{the:asymptotic}, one still needs to prove
that for any fixed value of $(Q_1^A,Q_1^B)\in\mathbb{R}_+^2$, the system formed by
equations~\Sinf admits at most one solution. The main ideas of this proof were evoked
in the proof of Proposition~1 of~\cite{partI}. 
However, the complete proof is omitted due to lack of space.


\begin{thebibliography}{25}

\bibitem{partI}
N. Ksairi, P. Bianchi, P. Ciblat and W. Hachem, \emph{Resource Allocation for
Downlink Sectorized Cellular OFDMA Systems: Part I---Optimal Allocation}, 2008. 


\bibitem{gau-hac-cib-1}
S. Gault and W. Hachem and P. Ciblat, \emph{Performance Analysis of an OFDMA Transmission System in a Multi-Cell Environment}, IEEE Transactions on Communications, num. 12, vol. 55, pp. 2143-2159, December, 2005.

%


%
%

\bibitem{asympOFDMA1}
J. Chen, R.A. Berry and M. L. Honig, \emph{Asymptotic Analysis of Downlink OFDMA Capacity},
Forty-Fourth Annual Allerton Conference, September 2006.

\bibitem{asympOFDMA2}
H. Wang and B. Chen, \emph{Asymptotic distributions and peak power analysis for uplink OFDMA signals},
IEEE International Conference on Acoustics, Speech, and Signal Processing, May 2004.

\bibitem{asympCDMA}
E. Biglieri, G. Caire, G. Tarrico and E. Viterbo, \emph{How fading affects
CDMA: an asymptotic analysis with linear receivers}, IEEE Journal on Selected
Areas in Communications, Volume 19, Issue 2, Pages:191 - 201, February 2001.

\bibitem{scaling_laws}
D. Gesbert, M. Kountouris, \emph{Joint Power Control and User Scheduling in
Multi-cell Wireless Networks: Capacity Scaling Laws,} submitted to IEEE Trans.
On Information Theory, September 2007.

\bibitem{bil-1}
P. Billingsley , \emph{Probability and Measure}, 3rd edition Wiley, New York,
1995.

\bibitem{sim_reuse}
M. Maqbool, M. Coupechoux and Ph. Godlewski,  
\emph{Comparison of Various Frequency Reuse Patterns for WiMAX Networks with
Adaptive Beamforming.} IEEE Vehicular Technology Conference, VTC, Singapore, may
2008.

\bibitem{sim_reuse2}
H. Jia, Z. Zhang, G. Yu, P. Cheng, and S. Li, \emph{On the Performance of IEEE 802.16 OFDMA System 
under Different Frequency Reuse and Subcarrier Permutation Patterns} Proc. of IEEE. ICC, June 2007.

\bibitem{sim_reuse3}
F. Wang, A. Ghosh, C. Sankaran, and S. Benes, \emph{WiMAX System
Performance with Multiple Transmit and Multiple Receive Antennas,}
Proc. of IEEE VTC, April 2007.

\bibitem{papandriopoulos}
T. Thanabalasingham,  S. V. Hanly,  L. H. Andrew and  J. Papandriopoulos,
\emph{ Joint Allocation of Subcarriers and Transmit Powers in a Multiuser OFDM Cellular Network},
IEEE International Conference on Communications ICC '06, June 2006. 

\bibitem{yates}
R. D. Yates, \emph{A Framework for Uplink Power Control in Cellular Radio Systems},
IEEE Journal on Selected Areas in Communications, vol. 13, no. 7, September 1995.

\bibitem{rapport}
N. Ksairi, P. Bianchi, P. Ciblat and W. Hachem, 
\emph{Resource Allocation for Downlink Cellular OFDMA Systems: Technical
Report}, 
Tech. Rep., May 2009. Available at 
http://www.tsi.enst.fr/$\sim$bianchi/Resource\_allocation\_technical\_report.pdf
.

\bibitem{perturbed}
J. F. Bonnans and A. Shapiro, \emph{Perturbation analysis of optimization
problems,} Springer, 2000.

\bibitem{okumura}
COST Action 231, \emph{Digital Mobile Radio toward Future Generation Systems,
final report}, Tech. Rep., European Communities, EUR 18957, 1999.

\end{thebibliography}
\end{document}